\documentclass[twocolumn,superscriptaddress,amsmath,amssymb,aps,pra,floatfix]{revtex4-1}

\usepackage[utf8]{inputenc}
\usepackage{wrapfig}
\usepackage{float}
\usepackage{amssymb}
\usepackage{amsfonts}
\usepackage{amsmath}
\usepackage{cases}
\usepackage{stmaryrd}\usepackage{tipa}
\usepackage[dvips]{graphicx}
\usepackage{dcolumn}
\usepackage{bbold}
\usepackage{graphicx}
\usepackage{graphicx}
\usepackage{subfigure}
\usepackage{dcolumn}
\usepackage{bm}
\usepackage{color}
\usepackage{xcolor}
\usepackage{CJK}
\usepackage{subfigure}
\usepackage{physics}
\usepackage{array}
\usepackage{multirow}
\usepackage{nopageno}
\usepackage{mathtools}
\usepackage{enumerate}
\usepackage{comment}
\usepackage{textcomp}
\usepackage{gensymb}
\usepackage[colorlinks=true, linkcolor=red, citecolor=blue]{hyperref}
\usepackage{siunitx}
\DeclareSIUnit{\belmilliwatt}{Bm}
\DeclareSIUnit{\dBm}{\deci\belmilliwatt}
\newcommand{\mathbbm}[1]{\text{\usefont{U}{bbm}{m}{n}#1}}
\definecolor{gg}{RGB}{8, 135, 68}
\definecolor{partner}{RGB}{255, 5, 228}
\newcommand{\Id}{\mathbbm{1}} 
 
\DeclareMathOperator{\diff}{d}

\begin{document}

\title{Propagating Quantum Microwaves: \\Towards  Applications in Communication and Sensing}


\author{Mateo Casariego}
\affiliation{Instituto de Telecomunicações, Physics of Information and Quantum Technologies Group, Portugal}
\affiliation{Instituto Superior Técnico, Universidade de Lisboa, Portugal}
\affiliation{Portuguese Quantum Institute, Portugal}
\author{Emmanuel Zambrini Cruzeiro}
\affiliation{Instituto de Telecomunicações, Physics of Information and Quantum Technologies Group, Portugal}
\author{Stefano Gherardini}
\affiliation{CNR-INO, Area Science Park, Basovizza, I-34149 Trieste, Italy}
\affiliation{Portuguese Quantum Institute, Portugal}
\author{Tasio Gonzalez-Raya}
\affiliation{Department of Physical Chemistry, University of the Basque Country UPV/EHU, Apartado 644, 48080 Bilbao, Spain}
\affiliation{EHU Quantum Center, University of the Basque Country UPV/EHU, Bilbao, Spain}
\author{Rui André}
\affiliation{Instituto de Telecomunicações, Physics of Information and Quantum Technologies Group, Portugal}
\author{Gonçalo Frazão}
\affiliation{Instituto de Telecomunicações, Physics of Information and Quantum Technologies Group, Portugal}
\affiliation{Instituto Superior Técnico, Universidade de Lisboa, Portugal}
\author{Giacomo Catto}
\affiliation{QCD Labs, QTF Centre of Excellence and InstituteQ, Department of Applied Physics, Aalto University, P.O. Box 13500, FIN-00076 Aalto, Finland.}
\author{Mikko Möttönen}
\affiliation{QCD Labs, QTF Centre of Excellence and InstituteQ, Department of Applied Physics, Aalto University, P.O. Box 13500, FIN-00076 Aalto, Finland.}
\affiliation{VTT Technical Research Centre of Finland Ltd. \& QTF Centre of Excellence, P.O. Box 1000, 02044 VTT, Finland.}
\author{Debopam Datta}
\affiliation{VTT Technical Research Centre of Finland Ltd. \& QTF Centre of Excellence, P.O. Box 1000, 02044 VTT, Finland.}
\author{Klaara Viisanen}
\affiliation{VTT Technical Research Centre of Finland Ltd. \& QTF Centre of Excellence, P.O. Box 1000, 02044 VTT, Finland.}
\author{Joonas Govenius}
\affiliation{VTT Technical Research Centre of Finland Ltd. \& QTF Centre of Excellence, P.O. Box 1000, 02044 VTT, Finland.}
\author{Mika Prunnila}
\affiliation{VTT Technical Research Centre of Finland Ltd. \& QTF Centre of Excellence, P.O. Box 1000, 02044 VTT, Finland.}
\author{Kimmo Tuominen}
\affiliation{Department of Physics and Helsinki Institute of Physics,  P.O.Box 64, FI-00014 University of Helsinki, Finland}
\author{Maximilian Reichert}
\affiliation{Department of Physical Chemistry, University of the Basque Country UPV/EHU, Apartado 644, 48080 Bilbao, Spain}
\affiliation{EHU Quantum Center, University of the Basque Country UPV/EHU, Bilbao, Spain}
\author{Michael Renger}
\affiliation{Walther-Mei{\ss}ner-Institut, Bayerische Akademie der Wissenschaften, 85748 Garching, Germany}
\affiliation{Physik-Department, Technische Universit\"{a}t M\"{u}nchen, 85748 Garching, Germany}
\author{Kirill G. Fedorov}
\affiliation{Walther-Mei{\ss}ner-Institut, Bayerische Akademie der Wissenschaften, 85748 Garching, Germany}
\affiliation{Physik-Department, Technische Universit\"{a}t M\"{u}nchen, 85748 Garching, Germany}
\author{Frank Deppe}
\affiliation{Walther-Mei{\ss}ner-Institut, Bayerische Akademie der Wissenschaften, 85748 Garching, Germany}
\affiliation{Physik-Department, Technische Universit\"{a}t M\"{u}nchen, 85748 Garching, Germany}
\affiliation{Munich Center for Quantum Science and Technology (MCQST), Schellingstr. 4, 80799 Munich, Germany}
\affiliation{IQM Germany GmbH, Nymphenburger Stra{\ss}e 86, 80335 M{\"u}nchen, Germany}
\author{Harriet van der Vliet}
\affiliation{Oxford Instruments NanoScience, Tubney Woods, Abingdon, Oxfordshire, OX13 5QX, UK}
\author{A. J. Matthews}
\affiliation{Oxford Instruments NanoScience, Tubney Woods, Abingdon, Oxfordshire, OX13 5QX, UK}

\author{Yolanda Fernández}
\affiliation{TTI Norte, 39011 Santander,  Spain}
\author{R. Assouly}
\affiliation{Ecole Normale Sup\'erieure de Lyon,  CNRS, Laboratoire de Physique, F-69342 Lyon, France}
\author{R. Dassonneville}
\affiliation{Ecole Normale Sup\'erieure de Lyon,  CNRS, Laboratoire de Physique, F-69342 Lyon, France}
\author{B. Huard}
\affiliation{Ecole Normale Sup\'erieure de Lyon,  CNRS, Laboratoire de Physique, F-69342 Lyon, France}

\author{Mikel Sanz}
\affiliation{Department of Physical Chemistry, University of the Basque Country UPV/EHU, Apartado 644, 48080 Bilbao, Spain}
\affiliation{EHU Quantum Center, University of the Basque Country UPV/EHU, Bilbao, Spain}
\affiliation{Basque Center for Applied Mathematics (BCAM), Alameda de Mazarredo 14, 48009 Bilbao, Basque Country, Spain}
\affiliation{IKERBASQUE, Basque Foundation for Science, Plaza Euskadi 5, 48009 Bilbao, Spain}
\author{Yasser Omar}
\affiliation{Instituto de Telecomunicações, Physics of Information and Quantum Technologies Group,  Portugal}
\affiliation{Instituto Superior Técnico, Universidade de Lisboa, Portugal}
\affiliation{Portuguese Quantum Institute, Portugal}

\begin{abstract}
The field of propagating quantum microwaves has started to receive considerable attention in the past few years. Motivated at first by the lack of an efficient microwave-to-optical platform that could solve the issue of secure communication between remote superconducting chips, current efforts are starting to reach other areas, from quantum communications to sensing. Here, we attempt at giving a state-of-the-art view of the two, pointing at some of the technical and theoretical challenges we need to address, and while providing some novel ideas and directions for future research. Hence, the goal of this paper is to provide a bigger picture, and -- we hope -- to inspire new ideas in quantum communications and sensing: from open-air microwave quantum key distribution to direct detection of
dark matter, we expect that the recent efforts and results in quantum microwaves will soon attract a wider audience, not only in the academic community, but also in an industrial environment.

\end{abstract}
\maketitle

\section{Introduction}

Despite the fact that photons at different frequencies are fundamentally the same thing, propagating quantum microwave technology is some 20 years behind the quantum optics at visible and infrared wavelengths. This is because up until recently we did not need quantum properties in a microwave field, although it is also true that the five orders of magnitude difference in energy have not made things easier: the way light interacts with matter naturally depends on its wavelength, and counting, or even detecting photons out of a field that does not trigger a photoelectric current is hard.

In classical domains such as radar, imaging, or mobile communication, microwave open-air technology is well established and omnipresent in everyday life. Hence, the question of how to extend such technology to the quantum regime is quite natural. In this context, one unexpected but nevertheless very important reason to study quantum microwaves comes from the field of quantum computing. There, one of the most promising platforms is based on superconducting circuits, which operate at microwave frequencies between \SIrange{1}{10}{\giga\hertz}. Superconducting quantum computing devices have reached a state of maturity where more than 100 coupled qubits can be controlled with high fidelity gates~\cite{IBM-Eagle}. Over the years, the requirement of cryogenic temperatures on the order of \SI{10}{\milli\kelvin} to preserve the quantum coherence of the qubits has been mitigated by robust commercial cryogenic technology. Just like in classical high-performance computing systems, the power of a quantum computer can be enhanced by further integration, and distributed via networked architectures. Regarding the latter, a microwave-to-optical transduction platform with high quantum efficiency would be desirable, but it is still quite out of reach of present-day technology. Hence, it is natural to consider propagating quantum microwaves for this task, because they intrinsically have zero frequency conversion loss, promising high remote-gate fidelities. This reasoning motivates the study of microwave quantum communications, a field that offers security protocols such as quantum key distribution (QKD) of paramount importance for the long-dreamed quantum internet~\cite{kimble2008}. Superconducting quantum chips communicating efficiently according to the laws of quantum mechanics represent the first major step in this direction: fully-microwave, operative quantum local area networks (QLANs).

Another motivation for using propagating microwaves in open-air without converting to optical comes from the fact that some of the current telecom infrastructure relies on these frequencies. The atmosphere has a transparency window for them, and their absorption is less impacted by unfavourable weather conditions than it is for so-called telecom frequencies. Combined, microwave and telecom links could further enable the reach of quantum communications.
For the same reason, the remote quantum sensing community has started to think about advantages of quantum microwaves in different metrology tasks. A prominent example here is the ongoing work towards a first demonstration of a quantum radar~\cite{macconeRadar, lanzagorta}, where the realisation of microwave quantum illumination in an open-air setting could represent a proof of principle experiment, although as we will argue, other approaches can be more practical in real-life scenarios.

Here, we give an overview of recent efforts to understand and tame propagating quantum microwaves, hinting at some promising directions along the way, and with an eye put on real-life applications for both communications and sensing.

The paper is organized as follows: in Section \ref{buildingBlocks} we review the state-of-the-art in quantum microwave technology: state generation, guided and non-guided propagation, Gaussian transformations, detection and state characterization techniques, signal amplifiers, and circulators. Section \ref{QCommunication} is devoted to the challenges in the field of quantum communications. We review some of the most recent results, such as quantum teleportation of an unknown microwave coherent state, and then move on to comment on the difficulties and benefits of proof-of-principle experiments such as inter-fridge QKD with microwave states.
Section \ref{QSensing} discusses quantum sensing, a field that we expect to benefit from recent developments in microwave photon-counting techniques. We discuss quantum illumination, quantum radar and imaging, and then mention two novel directions in quantum sensing: quantum thermometry, and direct detection of axionic dark matter.

\section{Building blocks of propagating quantum microwaves}\label{buildingBlocks}
The theory underlying propagating quantum microwaves is no other than quantum optics:
the quantized electromagnetic field is, in short, a continuous collection of frequency-dependent harmonic oscillators $\lbrace \hat{a}_\omega, \hat{a}^\dagger_\omega\rbrace_{\omega}$ with $\omega \in \mathbbm{R}^{+}$, where creation and annihilation operators satisfy the commutation relation $[\hat{a}_\omega, \hat{a}^\dagger_{\omega^\prime}] = \delta(\omega-\omega^\prime) \hat{\Id}$. Equivalently, and setting $\hbar = 1$, one can use two (dimensionless) orthogonal field quadratures for each mode $\omega$: $\hat{x}\equiv (\hat{a}+\hat{a}^\dagger)/\sqrt{2}$ and $\hat{p}\equiv (\hat{a}-\hat{a}^\dagger)/\sqrt{2}$. These quadratures are quantum continuous variables (CVs), in the sense that their expected values span the reals: $(x,p)\in \mathbbm{R}^2$ while satisfying the canonical commutation relation $[\hat{x}, \hat{p}] =i\hat{\Id}$. Additionally, photons carry a polarization degree of freedom, which can as well be used as a quantum information carrier. Often, the Wigner function --one of the quasi-probability distributions associated to a quantum state $\rho$, is used as an alternative description to the density operator. This function's operational interpretation lies closer to the CVs spirit, since its integral over some field quadrature is proportional to the probability of measuring the orthogonal one: $\int_\mathbbm{R} \diff p W(x,p) \propto \tr[\hat{x} \rho]$. This fact is used when performing Wigner-tomography, \textit{i.e.} reconstruction of a quantum state from the measured values of two orthogonal quadratures. Although in theory both quantum optics and quantum microwaves are described by the same formalism, \textit{e.g.} observables are obtained as some power series of $\hat{a}_\omega$ and $\hat{a}^\dagger_\omega$, the five orders of magnitude difference in energy makes the corresponding technology substantially different. In this section we review the most important steps and techniques related to the generation, transformation, and detection of propagating quantum microwaves, mentioning along the way some of the directions we expect to provide improvements in the short-term future of real-life applications in communications and sensing.

\subsection{State generation}\label{subsec:stategen}
Entanglement plays a central role in quantum technology. Typically bipartite, it arises when more than one mode of the electromagnetic field is considered. The word `mode' can mean different things, and essentially refers to the labels one uses to distinguish between Hilbert spaces: spatial modes give path-entanglement, frequency (or time of emission/arrival) modes give frequency-entanglement (or time-bins), polarization modes give discrete, polarization entanglement, and so on. All these types of entanglement are not necessarily mutually exclusive. Bipartite, mixed state entanglement, is commonly measured through the negativity, an entanglement monotone defined as $2\mathcal{N}(\rho):=\norm{\tilde{\rho}}_1-1$, where $\norm{\tilde{\rho}}_{1}:=\Tr\sqrt{\tilde{\rho}^\dagger \tilde{\rho}}$ is the trace norm of the partially transposed density operator.
Two-mode squeezed (TMS) states are natural quantum CVs candidates for communication and sensing protocols that require entanglement. They are routinely produced in labs. The most used devices to perform the squeezing operation in microwaves are Josephson parametric amplifiers (JPA) \cite{Eichler2011, Flurin2014}, which require a cryogenic environment to operate.  Symmetric two-mode squeezed states can be obtained either by directly pumping a parametric down-conversion term or by single-mode-squeezing two vacuum states -- which in practice are in thermal states -- in orthogonal directions, and then using a beam splitter to combine them. In either case, the resulting state is described by a two-mode squeezed thermal (TMST) state with mean photon number $N_\text{TMST} = 2n_\mathrm{th} \cosh(2r) + 2 \sinh^2(r)$, where $n_\mathrm{th}$ is the thermal photon number, and $r\in \mathbbm{R}$ is the squeezing parameter. For frequency-degenerate TMST states, a negativity $\mathcal{N}=3.9$ has been experimentally observed \cite{Fedorov:2018a}, with corresponding entangled-bit rate of $4.3\times 10^6 \text{ ebit}\cdot \text{s}^{-1}$, while in the frequency non-degenerate case, a rate of $6\times 10^6 \text{ ebit}\cdot \text{s}^{-1}$ has been reported \cite{Flurin2012}. 
Other approaches use traveling wave parametric amplifiers \cite{Esposito2021}, which is attractive for broad-band applications. Another way to generate TMS states consists in using a dc-biased Josephson junction in presence of two resonators. The entanglement production rate was more than $100\times 10^6 \text{ ebit}\cdot \text{s}^{-1}$ in Ref.~\cite{Peugeot2021}.

Cat states represent another source for CVs quantum entanglement with potential applications in microwave quantum technologies. In Ref.~\cite{Ma2019} these states are used to generate an entangled coherent state of two superconducting microwave resonators. 
Time-bin encoding in propagating microwaves has also been experimentally demonstrated \cite{Kurpiers2019}. These approaches are particularly interesting in scenarios where decoherence plays a role, since time bins are well known for their resilience against loss. Resilience of entanglement when these states propagate in open-air needs to be further studied in order to experimentally assess their utility for real-life applications.

Finally, polarization is a degree of freedom of  practical interest for propagating quantum microwaves. Current classical antennae designs already contemplate linearly or circularly polarized signals. However, microwaves in superconducting circuits have their polarization suppressed, as coplanar waveguides naturally imply a projection. The microwave-equivalent to an optical fibre, where the polarization vector could rotate freely, seems highly impractical due to the physical dimensions these fibres would need to have. Advances in 3D superconducting technology will be required to solve the issue \cite{Stammeier2017}.



\subsection{Propagation of quantum microwaves}\label{subsec:propagation}
\subsubsection{Guided}

Superconducting Niobium-Titanium coaxial cables are commonly used nowadays for low-loss guiding of microwave signals at cryogenic temperatures below $10$\,K. Such cables typically have a $50$\,$\Omega$ characteristic impedance and exhibit absorption losses on the order of $10^{-3}$\,dB/m for $\nu \simeq $\SI{5}{\giga \hertz}. These losses are mainly limited by the loss tangent of respective dielectrics, such as PTFE, and surface quality of superconducting material itself. Their diameter varies typically between $\sim$ 1\,mm and 3\,mm. In combination with crimped SMA connectors connectors, these cables offer a flexible, robust, and commercially available way to interface various devices in cryogenic environments. 

A somewhat alternative way for low-loss guiding of quantum microwave signals lies via using various rigid waveguides made of Niobium or Aluminum. Due to larger sizes of such waveguides, and therefore larger inner volumes, electromagnetic fields are strongly diluted in these systems, which reduces coupling to various dissipative channels. Additionally, these waveguides do not require the use of inner dielectrics. These factors lead to lower absorption losses below $5\cdot 10^{-4}$\,dB/m in the microwave waveguides. However, the price for this reduction comes in the form of their inflexible designs, connectors, and large sizes.

\subsubsection{Non-guided}

Non-guided propagation of microwave signals represents the typical scenario in open-air. This will be the result of transmitting a quantum state out of the cryostat by means of a quantum antenna. This device can be thought of as an inhomogeneous medium that smoothly connects two very different environments; a finite cavity that achieves impedance matching between the cryostat and the open-air. Let us briefly point out a very important difference between optical and microwave frequencies, noise-wise. Photons follow Bose-Einstein statistics: the walls of a cavity in thermal equilibrium at temperature $T$ are expected to produce an average photon number given by
\begin{equation}\label{eq:BEStats}
    n(\nu, T)= \frac{1}{e^{h\nu/k_B T}-1}
\end{equation}
 per unit volume at frequency $\nu$. At room temperature ($T=300$\,K), this gives an insignificant $n\sim  10^{-28}$--$10^{-55}$ in the optical domain (400-790\,THz), while a very noisy $n\sim 6250$--$625$ thermal microwave photons with $\nu$ in \SIrange{1}{10}{\giga \hertz}.

In a simple case study of a quantum antenna~\cite{GonzalezRaya2020}, the impedance function was optimized for the transmission of Gaussian microwave quantum states (two-mode squeezed states, in this case), from a cryostat at 50\,mK and associated impedance of 50\,$\Omega$, to open-air at 300\,K and associated impedance of 377\,$\Omega$. An exponential shape of the impedance, a well known result in the classical case, was found to reduce the reflectivity below $10^{-9}$. This showed that reducing losses in the antenna could lead both to the maximization of energy transmission, as well as to the preservation of quantum correlations.

Entanglement preservation was addressed by considering the antenna as a beamsplitter with a thermal noise input (using Eq.~\eqref{eq:BEStats} we find $ n(5\,\text{GHz}, 300\,\text{K}) \sim 1250$ photons), the latter being the main source of entanglement degradation. In a similar fashion, the reach of entanglement was studied in Ref.~\cite{GonzalezRaya2022}, considering an attenuation channel to describe absorption losses in a thermal environment. It was computed that entanglement could be transferred up to \SI{550}{\meter} in a realistic open-air setting through a two-mode squeezed thermal state generated at \SI{50}{\milli\kelvin} and with squeezing parameter $r=1$, considering oxygen molecules in the environment as the largest source of attenuation. Yet, a rigorous theoretical study of the atmospheric channel capacity \cite{Shapiro2005} in the microwave regime of interest for applications in communications and sensing is still missing to the best of our knowledge. 

\begin{figure}
\begin{center}
\includegraphics[width=0.46 \textwidth]{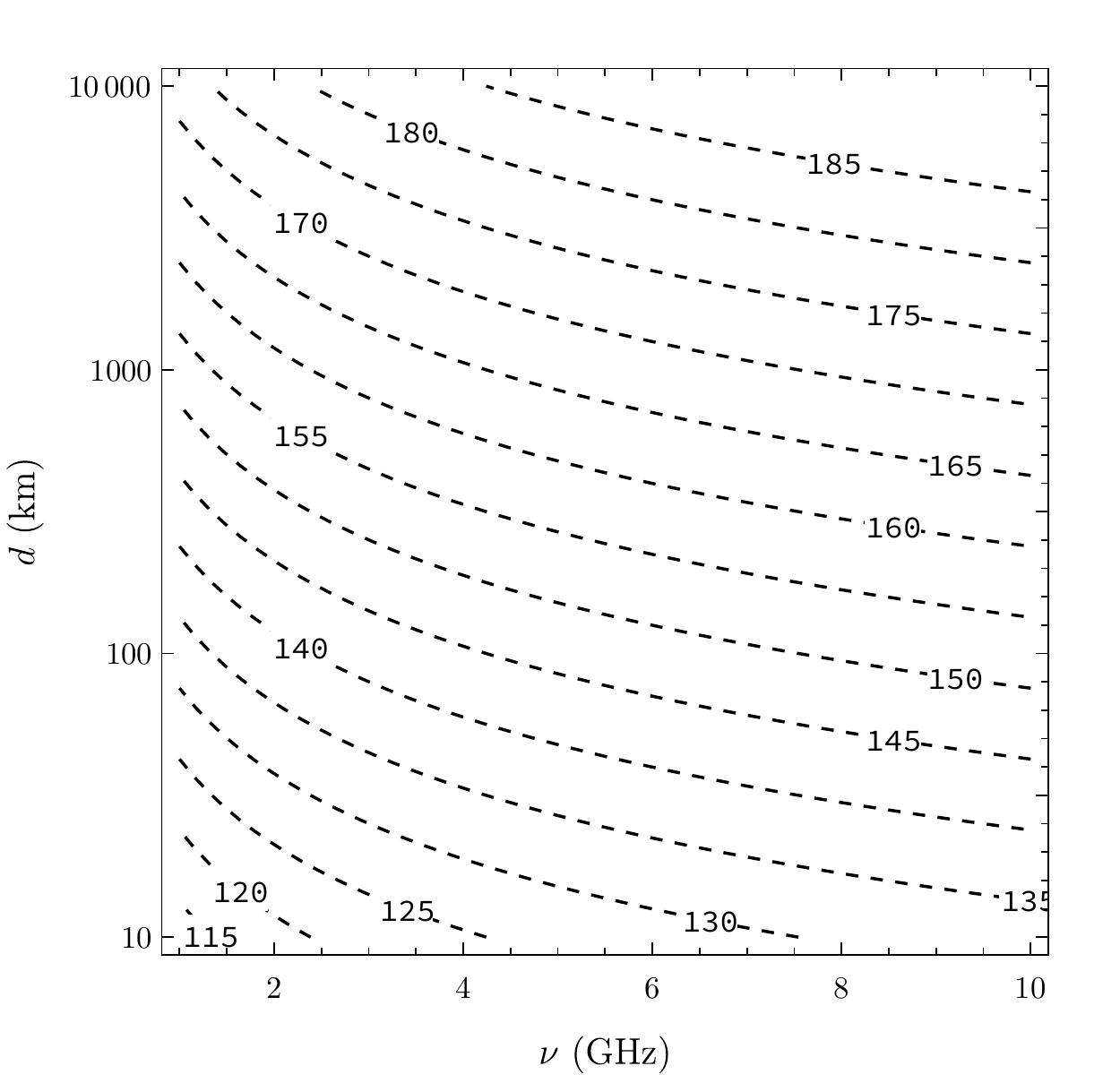}
\caption{Contour lines of the free-space path loss (FSPL) expressed in decibels as a function of the signal frequency $\nu$ (in the \SIrange{1}{10}{\giga\hertz} interval) and the distance $d$ between the emitter and the receiver in the range \SIrange{10}{10000}{\kilo\meter}: $L_\text{FSPL} (\text{dB})=20 \log_{10}\left(4\pi d\nu/c\right)$. FSPL quantifies a purely geometric  phenomenon arising from the natural three-dimensional spread of a signal in the far-field limit. The quadratic dependence of the FSPL on the frequency $\nu$, could make microwaves suitable for long-range broadcasting communications.}\label{fig:FSPLmw}
\end{center}
\end{figure}

Open-air and free-space propagation of quantum microwaves \cite{Pirandola2021, Pirandola2021-2, Kaltenbaek2021} requires addressing additional loss mechanisms: signal diffraction due to the natural spreading of an electromagnetic pulse in three dimensions; atmospheric attenuation; and wheather conditions. First, diffraction: We shall refer to this strictly geometrical loss as free-space path loss (FSPL). This is particularly relevant when large distances are considered, for example in inter-satellite or Earth-to-satellite links, or open-air communications on Earth. High Earth orbits (HEOs) are defined in the altitude range of 35786-$d_M/2$\, km, where $d_M$ is the distance from the Earth to the Moon. Space temperature in HEOs is roughly \SI{2.7}{\kelvin}, corresponding to the peak of the cosmic microwave background (CMB). These low temperatures could motivate the use of mechanically obtained cryogenics in satellites, as well as focalization lenses \cite{azad2017}. 
If an emitter and a receiver are separated a distance $d$ in a vacuum, with $d$ large enough so that the far-field approximation holds, the power ratio is given by Frii's transmission formula \cite{friis, hogg, shaw}:
\begin{equation}\label{eq:friis}
\frac{P_e}{P_r} =\frac{D_e D_d}{L_\text{A} L_{\text{FSPL}}},
\end{equation}
where $L_A$ is the absorption loss (that depends on wheather conditions), $L_{\text{FSPL}}=(4\pi d\nu/c )^2$ is  the geometric FSPL,  $\nu$ and $c$ are the frequency and the (vacuum) speed of the signal, respectively, and $D_{e,r }$ are the directivities of the emission and reception antennas. Directivity is defined as the maximized gain with respect to some preferred direction in space. It heavily depends on the design, ranging from no directivity at all (isotropic antenna with $D=1$) to high directional gains like the ones obtained with parabolic designs $D_\text{parabolic}= e_a (\pi a \nu /c)^2 $, where $a$ is the aperture, and $e_a$ is the aperture efficiency, a dimensionless parameter typically lying in \SIrange{0.6}{0.8}{} for commercial devices. Directivity gains are commonly given in $\text{dBi}=10\log_{10} D$. Importantly, the expression for $L_\text{FSPL}$ assumes a polarization-matching between emitter and receiver antennas. If this is not applicable, additional loss due to a polarization mismatch should be accounted for.  In Figure \ref{fig:FSPLmw} we plot the FSPL in dBs (i.e. the function $10 \log_{10}\left(L_{\text{FSPL}}\right)$) as a function of $\nu$ and $d$. 
Using Frii's formula alone as a comparison between optics and microwave signals is not completely fair, since it seems to indicate that microwaves are simply better for open-air communications. This is not necessarily the case. Traditionally, so-called `telecom' wavelengths are grouped in the following sets: near infra-red (NIR) with \SIrange{400}{207}{\tera\hertz}, short infra-red (SIR) with \SIrange{214}{100}{\tera\hertz}, mid infra-red (MIR) with \SIrange{100}{37}{\tera\hertz}, long infra-red (LIR) with \SIrange{37}{20}{\tera\hertz}, and far infra-red (FIR) with \SIrange{20}{0.3}{\tera\hertz} \cite{Kaushal2018}. In particular, the wavelengths \SIrange{780}{850}{\nano\meter} and \SIrange{1520}{1600}{\nano\meter} that belong to NIR and SIR ranges, respectively, enjoy of a very low atmospheric absorption loss of the order of \SI{0.1}{\dB \per \kilo\meter} in optimal wheather conditions. However, in the presence of dust, haze, or rain, Rayleight and/or Mie scattering needs to be accounted for. In Ref.~\cite{Fesquet2022} a full comparison between telecom and microwave regimes was made, concluding that the latter are more robust against unfavourable wheather conditions. In addition to this, the strong interaction of microwaves with non-linear elements give a higher entanglement rate production than telecom frequencies, which could justify the larger loss factor associated with the lack of highly directive emitters that is possible with telecom lasers.

\subsection{Gaussian transformations}
The manipulation of propagating electromagnetic signals via Gaussian transformations (50:50 beam splitting, displacement, phase rotation, squeezing) is well established at optical frequencies since many decades. There, many components used to guide and manipulate classical light are also known to work for quantum signals. However, despite the same theory description, the physical construction of components such as phase shifts, 50:50 beam splitters, displacers, or squeezers is different in the microwave regime of \SIrange{1}{10}{\giga\hertz}, because, there, the wavelength is roughly five orders of magnitude larger, i.e., on the order of a few centimeters.

When the research community started to think about the quantum properties of propagating microwaves in the late 2000s, first linear operations had been addressed. Due to the convenient wavelength, phase shifts can be achieved rather trivially via short pieces of delay line. The construction of microwave beam splitters and displacers is more involved; they are interference devices with typical dimensions on the order of the operation wavelength. When respecting the relevant boundary conditions of unitarity (absence of loss), impedance matching, and isolation between ports, a 50:50 beam splitter is always a four-port device. In terms of quantum mechanics, these conditions ensure energy conservation and commutation relations. Experimentally, vacuum fluctuations as fundamental  as the minimal noise added to the input signal of a 50:50 beam splitter were first discussed in Ref.~\cite{Mariantoni:2010a}. Interestingly, even devices with only three signal connectors were shown to have a fourth internal port there. 50:50 beam splitters are used to create superpositions or path entanglement between two microwave beams~\cite{Menzel:2012a}, in microwave interferometers~\cite{Eder:2018a}, for dual-path tomography~\cite{Menzel:2010a,dicandia2014}, and in the feedforward mechanism of quantum teleportation~\cite{Fedorov:2021a}. In these experiments, often commercial conducting devices (\textit{i.e.} non-superconducting) with little dissipation on the order of \SI{0.3}{\deci\bel} are enough. For more delicate situations where even little dissipation is harmful, superconducting beam splitters were developed~\cite{Hoffmann:2010a,Ku:2011a,Eder:2018a}.

The second important linear Gaussian transformation is the displacement operation $\hat{D}=\exp(\alpha\hat{a}^\dag-\alpha^*\hat{a})$. The name of this transformation is inspired by the fact that it actually displaces the Wigner function of the state of a bosonic mode $\hat{a}$ in phase space by the vector $\alpha$. Experimentally, the displacement is implemented by weakly coupling a coherent drive of complex amplitude $\alpha$ to the propagating mode $\hat{a}$ via a strongly asymmetric (typically 99:1) beam splitter. The corresponding microwave device is commonly called a directional coupler and, again, exists in low-loss normal-conducting commercial~\cite{Fedorov:2016a}, and laboratory-made superconducting~\cite{Ku:2011a} variants. Since its demonstration for propagating microwaves~\cite{Fedorov:2016a}, displacement is being routinely used in the feedforward mechanism of microwave continuous-variable quantum communication protocols~\cite{Pogorzalek:2019a,Fedorov:2021a}.

\begin{figure}
\begin{center}
\includegraphics[width=0.4 \textwidth]{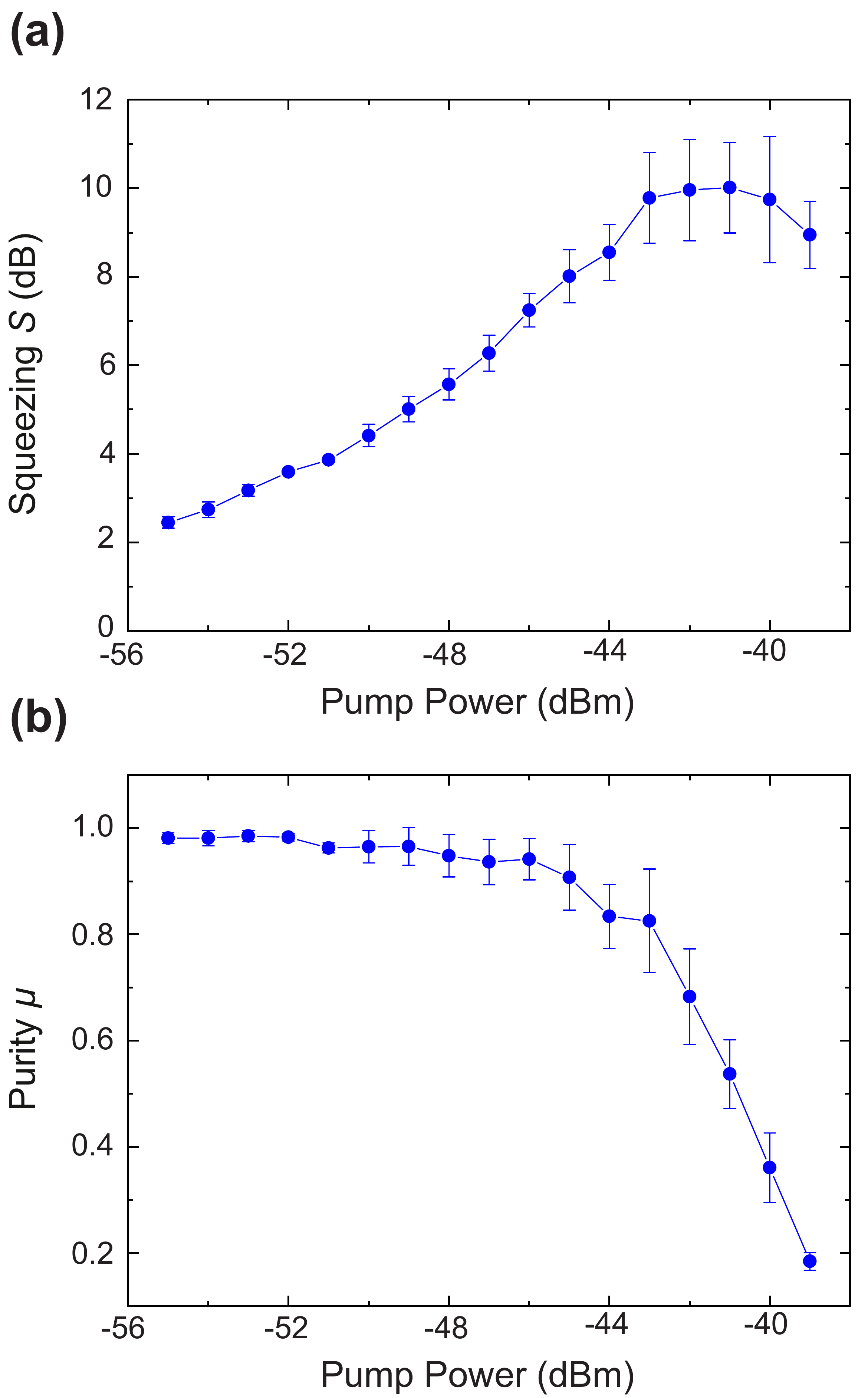}
\caption{(\textbf{a}) Squeezing as a function of pump power for a single-SQUID niobium JPA. The pump power is referred to the JPA input. (\textbf{b}) Corresponding purity of the reconstructed squeezed states as a function of the pump power.}
\label{fig:SqueezingPurity}
\end{center}
\end{figure}

The generic nonlinear Gaussian transformation for a single mode is called squeezing, because one quadrature variance of a Gaussian state is reduced below the corresponding variance of the vacuum state. In turn, the orthogonal quadrature must be enlarged to respect the Heisenberg principle. In a Wigner function picture, the name can be intuitively understood because the circular blob of the vacuum state is squeezed into a narrow elliptic shape. The single-mode squeeze operator $\hat{S}=\exp\left(\frac{1}{2}\xi^*\hat{a}^2-\frac{1}{2}\xi\left(\hat{a}^\dag\right)^2\right)$ describes the squeezing operation on the mode $\hat{a}$. Here, the complex parameter $\xi$ is related to the squeezing parameter $r$ via $\xi = r e^{i\theta}$, and controls the suppression (``squeezing'') and enlargement (``anti-squeezing'') of two orthogonal field quadratures, as well as the orientation of the squeezed quadrature in phase space. In general, squeezing is created via a parametric process, where a system parameter is modulated with a suitable high-frequency signal. Squeezed microwave signals are commonly generated from a superconducting LC circuit, where part of the inductance is a current- or flux-tunable Josephson inductance. The parametrically induced nonlinearity can be conveniently amplified in narrowband resonant devices such as Josephson parametric amplifiers (JPA)~\cite{Zhong:2013a} or Josephson parametric converters (JPC)~\cite{Bergeal:2010a}. When broadband operation is desired, a successful solution has been to use an open transmission line with many Josephson devices and a suitable phase matching. These devices, which are known as Josephson travelling-wave parametric amplifiers (JTWPAs)~\cite{Perelshtein:2021a,Macklin:2015a,White:2015a}, are also important for the high-fidelity single-shot qubit readout in superconducting quantum computing architectures. Most experiments based on flux-driven JPAs~\cite{Zhong:2013a} employ propagating microwave modes with up to \SI{9}{\deci\bel} of squeezing~\cite{Fedorov:2018a}. However, for squeezing beyond \SI{3}{\deci\bel}, induced noise results in a significant thermal contribution and, hence, a significantly reduced purity of the squeezed state~\cite{Pogorzalek:2019a}. Recent experiments on the amplification properties of JPAs suggest pump-induced noise as a limiting factor for quantum efficiencies~\cite{Renger:2021a}. Figure\,\ref{fig:SqueezingPurity} shows an exemplary plot for the pump power dependence of reconstructed single-mode squeezing (\textbf{a}) and purity (\textbf{b}) for a single-SQUID niobium JPA, fabricated by VTT. Here, we define the squeezing level as $S = -10 \log(\sigma_\mathrm{s}^2/0.25)$, where $\sigma_\mathrm{s}^2$ denotes the squeezed variance. The purity has been calculated from the corresponding reconstructed covariance matrix $V$ by $\mu = 1/(4\sqrt{\det V})$.

By means of a symmetric beam splitting operation, the quantum correlations inherent to squeezed states can be transferred to entanglement correlations -- also called two-mode squeezing -- between the split beams. For frequency-degenerate two-mode squeezing, the two operations are usually executed in series~\cite{Menzel:2012a,Fedorov:2018a}. Here, the entanglement must be distributed among two distinct physical paths (``path entanglement''). State-of-the-art devices exhibit an entanglement of formation on the order of a few Mebits/s over the full JPA bandwidth, which would be usable for quantum communication purposes~\cite{Fedorov:2018a}. Note that an ``ebit'' is a logical unit of bipartite entanglement. A maximally-entangled pair of qbits, for example in a Bell state, carry one ebit. For frequency-nondegenerate two-mode squeezing, both operations can be combined into a single device such as the above-mentioned ~JTWPA or a JPC. In the former case, the two-mode squeezing coexists in the same beam at two different frequencies. For a JPC, the entanglement is between two different beams at two different frequencies.

Recently, single-mode and two-mode squeezing have been obtained not only with narrowband superconducting parametric devices, but also with a JTWPA. Operating in dual-pumped, non-degenerate four-wave mixing, the device provides \SI{-11.35}{\deci\bel} of single-mode squeezing and an average of \SI{-6.71}{\deci\bel} of two-mode squeezing over a bandwidth of \SI{1.75}{\giga\hertz}.
The temperature regime of these experiments is on the order of \SI{10}{\milli\kelvin}, to avoid thermal fluctuations at the input of the device.

Finally, another important device class for experiments with propagating microwaves are circulators and isolators, where the latter can be described by a circulator with one \SI{50}{\ohm}-terminated port. A circulator is an $n$-port device with a time-reversal symmetry breaking mechanism. As a consequence, from each port, an incoming signal can only propagate towards one of the two neighboring ports, thereby forming a directive element. In other words, if inputs are called $I_k$ and outputs $O_k$, with $k\in [1,\ldots, n]$, then a circulator satisfies $I_k = O_{k+1}$. Circulators are of eminent practical importance for the currently known communication and sensing protocols with propagating microwaves. In a cryogenic environment, passive circulators based on ferrite materials biased by permanent magnets are widely used because of their favorable combination of sufficient bandwidth, insertion loss, isolation, input power tolerance, and commercial availability. Their main drawbacks are the requirement of significant magnetic bias fields, incompatibility with on-chip integration into superconducting circuits, and bulkiness due to the interference concept and magnetic shielding requirements. Furthermore, even the small insertion loss of \SI{0.3}{\deci\bel} from their normal conducting microwave circuitry adds up when using multiple circulators in series in more complex experimental settings. As a way out, circulators based on Josephson or nanomechanical elements have been proposed and implemented~\cite{Chapman:2017a,Ranzani:2019a}. Although these devices can be on-chip integrated with other superconducting quantum circuits and have potential to operate at the quantum limit, they still suffer from complex fabrication, demanding flux tuning, or low input power tolerance.

\subsection{Detection of quantum microwaves}\label{subsec:detection}
State characterization and quantum tomography techniques typically rely on extracting information from a large sample of states generated under the same conditions. The way this information is extracted depends on the detection techniques one has available, and this naturally varies depending on the frequency of the field under observation. 
Quantum sensing and communication protocols requiring feedforward control are more challenging, as a high signal-to-noise ratio and fast signal processing are required for making a decision in a single shot, i.e., without ensemble averaging. 
In this section, we review the state-of-the art in detection of quantum microwaves. This also involves the step of amplifying, and, or downconverting the signals. The results discussed here are all related to cryogenic settings. The problem of open-air detection at higher temperatures still remains a technological challenge. 
\subsubsection{Homodyne, and heterodyne detection}
Photodetectors, not to be confused with photon-counters, are devices capable of transforming a photocurrent $\hat{i} \propto \hat{a}^\dagger \hat{a}$ into an electric current. In microwaves, photodetection has been experimentally demonstrated in a variety of settings  \cite{milford2006, romero2009, chen2011, peropadre2011, sathyamoorthy2016, Govenius2016}.
Homodyne detection relies on this concept in order to extract information about a single quadrature of the electromagnetic field. Heterodyne detection, on the other hand, measures two orthogonal quadratures at the same time, allowing a complete quantum state reconstruction. Most quantum microwave tomography techniques use some form of heterodyne measurements \cite{Eichler2011, menzel2012}.
In Ref. \cite{eichler2012}  a single-shot heterodyne detection scheme is proposed for microwaves, which makes use of a linear phase-insensitive cryogenic amplifier, a mixer, and analog-to-digital converters (ADC). State tomography for single Fock itinerant microwave states has been experimentally demonstrated with linear amplifiers and ADCs \cite{Eichler2011-2}. Using an NV-center as mixer, in Ref.~\cite{Meinel2021} a heterodyne measurement scheme with spectral resolution below \SI{1}{\mega \hertz} for a \SI{4}{\giga \hertz} signal is proposed.
 Closely related to heterodyne detection are CVs Bell measurements. In Ref.~\cite{Fedorov:2021a} these are realized using two phase-sensitive amplifiers together with two hybrid rings, and a directional coupler. Bell measurements are required in most implementations of CVs quantum teleportation, as well as in situations where one needs to guarantee a loophole-free Bell test scenario.
 
It is expected that advances both in microwave photon-counting and amplifiers will improve the quality of single-shot homodyne measurements, and consequently of Bell measurements.

\subsubsection{Qubit-based photon counting}
Many scenarios in quantum communication, sensing, or computation,  require the use of non-Gaussian operations and measurements. The most notable one is photon counting. As an example, as described below, quantum illumination requires photon counting to obtain a quantum advantage. Photodetectors can also be used to implement the full state tomography of a propagating quantum state~\cite{besse2020}. And to distribute entanglement between remote qubits using a joint measurement~\cite{Narla2016}. Furthermore, photodetectors allow one to build heralded noiseless amplifiers, which can restore lost entanglement in dissipative transmission lines or owing to imperfect generation~\cite{Xiang2010}.

First, it is instructive to distinguishing counters of stationary versus propagating modes. The first kind has been demonstrated since 2007 with Rydberg atoms and with superconducting circuits~\cite{Gleyzes2007,Guerlin2007,Johnson2010,Leek2010,Sun2014} and is now a standard component of the circuit-QED toolbox. In contrast, the detection of propagating microwave photons is more challenging and many implementations have been proposed~\cite{Romero2009,Helmer2009,Koshino2013,Sathyamoorthy2014,Fan2014,Kyriienko2016,Sathyamoorthy2016,Gu2017,Wong2017,Leppakangas2018,Royer2018}. First experimental realizations used direct photon absorption by a Josephson junction~\cite{Chen2011,Inomata2016}, which leads to a destructive photodetector. More recently, quantum non-demolition detectors have been demonstrated either by encoding the parity in the phase of a qubit~\cite{Besse2018,Kono2018} or by encoding the presence of a single photon in a qubit excitation~\cite{Narla2016}, also combined with reservoir engineering~\cite{Lescanne2019}. 

Until recently, and despite several proposals of a photocounter -- a microwave photodetector able to resolve the photon number -- for a propagating mode~\cite{Romero2009,Royer2018,Kono2018,Sokolov2020,grimsmo2020quantum}, an experimental realization was missing. In Ref.~\cite{Dassonneville2020}, a photon-counter working within a \SI{20}{\mega\hertz} band centered around \SI{10.220}{\giga\hertz} is able to distinguish between 0, 1, 2 and~3 photons, with detection efficiencies of $99\%$ for no photons, $(76\pm3)\%$ for a single photon, $(71\pm3)\%$ for two photons, and $(54\pm2)\%$ for three, with a dark count probability of $(3\pm0.2)\%$, and an average dead time of \SI{4.5}{\micro\second}. Microwave photon detectors and counters enable a variety of applications. In Ref.~\cite{Balembois2021}, a state-of-the-art photodetector has been used to improve the sensitivity of electron spin resonance detection. 

\subsubsection{Bolometers for photon counting}\label{subsubsec:bolometers}
A bolometer is a device that detects heat deposited by absorbed radiation and converts it into an electric signal. In contrast to qubit detectors, a bolometer typically has a resistive input, and hence can operate on a broad input frequency band. A bolometer that has a fast readout and low enough noise, can be used as a calorimeter, i.e., to measure the energy of an incoming wave packet. If the frequency of the input photons is known, such a calorimeter also works as a photocounter. In addition, a bolometer is typically read out continuously with no dead time. However, the challenge in using bolometers in the single-photon microwave experiments has been their too high noise and slow thermal time constant, which has been greatly relieved thanks to recent developments~\cite{Kokkoniemi2020}. This motivates us to discuss the recent advancements in bolometers.

The most promising bolometers for cQED consist of a two-dimensional graphene flake~\cite{Kokkoniemi2020, Lee2020} or a metallic gold-palladium nanowire~\cite{lowNEP} connected to a temperature-dependent effective inductance implemented using the superconductor proximity effect~\cite{govenius2014microwave, proximity2017}. Changes in the inductance change the resonance frequency of an $LC$ tank circuit and hence the reflection coefficient of the roughly \SI{600}{\mega\hertz} probe signal. Thus, by continuously digitizing the probe signal, one can continuously monitor the temperature or the absorbed power at the bolometer input.

Owing to the electrothermal feedback, i.e., the probe tone also heating the bolometer depending on the resonance frequency, the $LC$ tank circuit exhibits a bistable regime~\cite{zeptoj} at a certain range of input powers and probe frequencies. Thus, by fine-tuning the probe parameters we can optimize for high signal-to-noise ratio or speed of thermal relaxation. The optimal operation point typically lies close but outside the regime of bistability. 

The noise equivalent power (NEP) is a typical figure of merit for bolometers and equals to the noise spectral density in the bolometer readout signal in the units of the input power to the bolometer. The recently discovered metallic bolometers have showed an NEP as low as $20$~zW/$\sqrt{\text{Hz}}$ \cite{lowNEP} with a thermal time constant of $30$~ms, leading to an extracted bolometer energy resolution of $h\times400$~GHz. On one hand, this energy resolution and speed are not satisfactory for single-photon microwave counting, but on the other hand, they are not many orders of magnitude off. Fortunately, using graphene as the proximity superconductor, in order to lower the bolometer heat capacity, and hence to maximize the induced temperature change due to an absorbed photon, the obtained results are further improved: with a similar NEP of $30$~zW/$\sqrt{\text{Hz}}$, and a much improved time constant in the hundred-nanosecond scale, and an extracted energy resolution as low as $h\times30$~GHz~\cite{Kokkoniemi2020}. Although the energy resolution is still not meeting the requirements of photon counting for typical frequencies in cQED, $\lesssim10$~GHz, these numbers seem satisfactory to start using the bolometer as a readout device for superconducting qubits, where the information of the qubit state is encoded into a microwave pulse of several photons.

The above-described bolometer is a robust and easy to operate device. Its footprint on the chip is not larger than that of typical transmon qubits and it is capable of detecting absorbed photons in a broad bandwidth in real time. Neither photon shaping nor knowledge of its arrival time is required. Thus, some further optimization of the graphene bolometer may lead to a convenient device for photon counting for the cQED. In addition to its wide range of applications in photon sensing in general, this device can be modified to detect the heat deposited by a dc current, also allowing one to calibrate for absolute microwave power or possibly low currents at millikelvin temperatures~\cite{girard2021cryogenic}.

\subsubsection{Low-noise cryogenic amplifiers}
In most applications, readout signals from the quantum circuits need to be eventually read out and processed at room temperature. Thus, a low-noise and high-gain amplification chain is required to amplify the weak signals originating from quantum circuits, typically located at millikelvin temperatures, before they pass through room-temperature components.
Typical output powers originating from quantum circuits range between roughly \SI{-150}{\dBm} for single-photon signals and  \SI{-120}{\dBm} used in dispersive qubit readout.
The first stage of amplification is usually provided by superconducting parametric amplifiers, as they have been demonstrated to provide near-quantum-limited noise performance as well as sufficient bandwidth and power handling for most applications. The next-stage amplification is typically provided by low-noise cryogenic high-electron-mobility transistor (HEMT) amplifiers at around \SIrange{3}{4}{\kelvin}, followed by a final post-amplification stage at room temperature.
The gain and noise temperature of each amplification stage is usually chosen such that system noise, calculated using Friis formula, remains close to the noise added by the first amplifier.

Regarding the cryogenic low-noise solid-state amplifier technologies, the lowest noise temperatures are achieved based on Indium-Phosphide (InP) HEMTs. In Ref.~\cite{Schleeh2012} a \SIrange{4}{8}{\giga\hertz}, three-stage, hybrid low-noise amplifier operating at \SI{10}{\kelvin} is proposed, providing an average noise temperature of \SI{1.6}{\kelvin}. The gain of the amplifier across the entire band is \SI{44}{\deci\bel}, consuming \SI{4.2}{\milli\watt} of DC power. Future quantum computing applications demand a number of readout channels, integrated within a single cryogenic system, of the order of $10^3$. Improvements in power consumption will be key to efficiently enable future quantum technologies.  Recently, an ultra-low power \SIrange{4}{8}{\giga\hertz} InP HEMT cryogenic low-noise amplifier (LNA) has been demonstrated \cite{Cha2020}, achieving an average noise of \SI{3.2}{\kelvin} with \SI{23}{\deci\bel} gain, and an ultra-low power consumption of just \SI{300}{\micro\watt}. Apart from InP HEMTs, Silicon-Germanium (SiGe) heterojunction bipolar transistors (HBTs) are appearing as promising cryogenic LNAs candidates because they are compatible with complementary metal–oxide–semiconductor (CMOS) technology, making them more appropriate for uses in large-scale systems. A \SIrange{4}{8}{\giga\hertz} SiGe cryogenic LNA has been implemented using the BiCMOS8HP process \cite{Montazeri2017}, providing \SI{26}{\deci\bel} of gain while dissipating \SI{580}{\micro\watt} of DC power. The noise temperature was \SI{8}{\kelvin} across the frequency band, so significant research is still required to optimize the cryogenic SiGe HBTs performance.

At cryogenic temperatures superconducting parametric amplifiers, such as JPAs, play an important role for microwave quantum technology. These devices exhibit ultra-low power dissipation in the relevant temperature range due to their superconducting properties. Moreover, these amplifiers can be operated close the fundamental quantum amplification limit of $1/2$ added noise photons, also known as the standard quantum limit (SQL) of phase-insensitive amplification. Fundamentally, the SQL originates from the commutation relation of bosonic operators describing electromagnetic fields in quantum mechanics. It should be noted that in the framework of QMiCS, sub-GHz JPAs have also been used to improve the performance of the microwave calorimeters described in Sect. \ref{subsubsec:bolometers}  (see also Ref. \cite{lowNEP}). 

Resonator-based JPAs typically suffer from a limited gain-bandwidth product, often on the order of \SIrange{0.1}{5}{\giga\hertz}. This constraint limits their applications in many practically relevant applications, such as broadband frequency-multiplexed qubit readout or generation of cluster states, among others. Furthermore, conventional superconducting JPAs used to suffer from limited \SI{1}{\deci\bel} compression point, on the  order of \SI{-120}{\dBm}. In recent years, it has been shown that the latter limitation can be circumvented by using multiple superconducting nonlinear elements, such as SQUIDs or SNAILs \cite{Frattini:2017}, in combination with the CPW resonators. Alternatively, one can exploit superconducting materials with high kinetic inductance, which can push the \SI{1}{\deci\bel} compression point as high as \SI{-50}{\dBm} for typical microwave frequencies around \SI{5}{\giga\hertz}.

In applications requiring higher bandwidth, superconducting traveling-wave amplifiers are used. Although parametric amplifiers were originally proposed even before the transistor amplifier, the recent rise of popularity has happened in last decade. 
JTWPAs introduced in \ref{subsec:Gaussian transformations} consists of array of Josephson junction-based circuit elements\cite{Macklin:2015a,Miano:2019,Perelshtein:2021a,Frattini:2017} fabricated on a superconducting transmission line which acts as a 'non-linear media' facilitating wave mixing and amplification. Similarly, other efforts have fabricated the non-linear media using high-kinetic inductance material \cite{Malnou:2021}, which facilitates non-linear inductance. The later devices are often named as Kinetic Inductance TWPA (KI-TWPA). In JTWPAs, in addition to the Josephson elements various dispersion engineering or frequency domain band engineering are implemented in order to achieve phase matching \cite{Esposito:2021a}. Whereas in KI-TWPA such dispersion engineering are achieved by periodically loading the transmission line with capacitive elements. 
Early realization of JTWPAs operating in degenerate four-wave mixing mode (4WM) have produced gain in excess of \SI{15}{\deci\bel} and noise performance close to standard quantum limit \cite{Macklin:2015a}, with a disadvantage that the pump signal resides in the same frequency band as the amplified quantum signals, making filtering cumbersome and risking saturation of subsequent components in the readout chain. In contrast, three-wave mixing (3WM) TWPAs by device design have the advantage of the pump signal being situated outside the signal band, facilitating easy filtering \cite{Ranadive:2021,Perelshtein:2021a}.
Recently, a non-degenerate 4WM TWPA with two pump tones has been demonstrated \cite{Jack Y.Qiu:2022}, providing multiple GHz of signal bandwidth between the pump tones without sacrificing gain and noise performance. 
Similarly, 3WM implementations of both JTWPAs \cite{Perelshtein:2021a} and KITWPAs \cite{Malnou:2021} have produced gain in excess of \SI{15}{\deci\bel} in the \SIrange{4}{8}{\giga \hertz} frequency range with near-quantum-limited noise performance in the same frequency band.
Although the gain performance as a first stage amplifier in readout chain provided by both technology platform are comparable, KITWPAs provide significantly larger \SI{1}{\deci\bel}-compression power of roughly \SI{-60}{\dBm} \cite{Malnou:2021}, which is almost \SI{20}{\deci\bel} higher compared to that of state-of-the-art JTWPAs. The higher compression power provides advantages in multiplexed read out of detectors, sensors and other devices.  
Satisfactory gain and noise performance has been achieved with broadband superconducting TWPAs, in view of Friis formula for system noise. However, compared to conventional semiconductor amplifiers, TWPAs still suffer from relatively high ripple in the signal band, generation of spurious frequency tones, less standardized fabrication, and a delicate tune up procedure. Furthermore, superconducting TWPA performance is strongly dependent on the quality of broadband matching provided by RF components at the input and output of the JTWPA.Hence, future improvement of the superconducting parametric amplifier as a first-stage amplifier platform would require improvement in associated microwave components.    

\section{Quantum Communication}\label{QCommunication}

Quantum communication promises unconditional security in private communications, as long as nature behaves in accordance with the laws of quantum physics. This has been recently demonstrated in proof-of-principle experiments~\cite{Nadlinger2021,Zhang2021,Liu2021}. Such remarkable proof-of-principle experiments became feasible thanks to many developments such as the experiments that closed the locality and detection loopholes in 2015~\cite{Hensen2015,Shalm2015,Giustina2015}, theoretical advances in understanding the certification of quantum correlations~\cite{Arnon2018,Nadlinger2021}, and major experimental achievements in linear optics~\cite{Liu2021} and light-matter interaction for example with trapped ions~\cite{Stephenson2020,Nadlinger2021}. Quantum communication has clearly come a long way since its inception in 1984 with the celebrated paper by C. Bennett and G. Brassard~\cite{Bennett84}, describing the first quantum key distribution protocol: BB84. 

Since then, a panoply of quantum communication protocols have been developed, employing different types of degrees of freedom of particles, such as position in space, time, frequency, polarization, and spin. For example, with free-space linear optics one often uses polarization for its simplicity and availability of polarization optics. Nevertheless in fiber networks, where polarization losses can be important, a degree of freedom such as time is much more convenient. More generally, one may distinguish discrete degrees of freedom from continuous ones, giving rise to DV-QKD and CV-QKD, respectively. 

A long list of protocols have been proposed for DV-QKD, some of which are: E91~\cite{Ekert1991}, B92~\cite{Bennett1992}, COW~\cite{Stucki2005}, variations of BB84 such as the ones using decoy states~\cite{Lo2005,Grunenfelder2018}. 
Although CV-QKD is a more recent topic, there exist many protocols, making use of various quantum resources, i.e., states and measurements. On the one hand, security proofs are harder to obtain with respect to the DV case because of the need to work with infinite dimensional states and unbounded measurement operators. On the other hand, the technology for CV QKD, i.e., coherent detection, is more readily available. Both approaches are therefore very promising.

Finally, note that now quantum communication has been taken to space, with spectacular experiments such as 1200 km quantum teleportation~\cite{Ren2017,Liao2017}, and entanglement-based QKD demonstration over the same distance~\cite{Yin2020}.

Although quantum communication experiments have seen an impressive flourishing over the past four decades thanks to a better understanding of light and light-matter interaction, the idea of microwave quantum communication has not seen such a dramatic development. In fact, many questions remain: What is the feasibility of CV and DV quantum communication protocols with microwaves? How does one efficiently transmit a quantum state to open-air, given the dramatic differences in temperature and impedance between the two environments? Can microwave quantum communication be easier implemented in some situations? For example, can it be advantageous for satellite communication? 
Is it better to communicate with microwave or optical photons in a turbulent/polluted atmosphere? 
Some of these questions have been partly answered already in Section \ref{subsec:propagation}. In this section, we will review the state of the art, and explore what is being done currently to accelerate the development of microwave quantum communication. 

The section is organized as follows. First, in Section~\ref{subsec:swap} we show experiments that generate and manipulate quantum resources, such as entanglement swapping and teleportation experiments. Then, in Section~\ref{subsec:qkd} we present some QKD experiments, while in Section~\ref{subsec:qLAN} quantum local area networks and the scaling to a quantum internet are finally discussed.

\subsection{Distribution of quantum resources}
\label{subsec:swap}

As discussed in Section \ref{subsec:stategen} and illustrated with the example of linear polarization, the generation of quantum resources for CV QKD is more straightforward than for DV QKD. For this reason, it would be of great value to develop techniques to generate DV quantum states, not only in polarization, but also in other degrees of freedom which could be easier to implement, as for example it has been already demonstrated for time-bins \cite{Kurpiers2019}.


The first microwave quantum teleportation demonstrations were performed in an intra-fridge setting, usually over distances shorter than $1$ meter. Quantum teleportation in the microwave regime has been performed both with DV \cite{Wallraff2014} and CV \cite{diCandia2015,Fedorov:2021a}. Intra-fridge quantum state transfer has been also demonstrated \cite{Kurpiers2018}. 

In the inter-fridge scenario, quantum states can be transferred from a cryostat to another one using a microwave quantum link both with DV and CV. In the DV case (transmon qubits), one can both prepare quantum states and measure them on the other end of the channel, and share entanglement (Bell states) between both ends, with fidelities around $85.8\%$ and $79.5\%$ respectively \cite{Magnard2020}. In the CV case, the experiment has not been performed yet.

Recently, a study about the feasibility of open-air microwave entanglement distribution for quantum teleportation of CVs was presented  \cite{GonzalezRaya2022}. There, absorption losses in a thermal environment are taken into account, to obtain an upper bound of \SI{550}{\meter} for the maximum distance that a TMST state generated at \SI{50}{\milli\kelvin} with squeezing parameter $r=1$ can propagate before completely losing its entanglement. To overcome this, non-Gaussian subroutines like photon-subtraction are proposed, arguing that these that could enhance significantly the degree of two-mode CV entanglement and therefore the overall quality of the protocol.

\subsection{Quantum key distribution}
\label{subsec:qkd}

Recently, theoretical studies taking into account the recent developments in propagating microwaves have indicated the significant potential brought by quantum microwaves for both short distance quantum communication \cite{Diamanti2016}, and long distance in the context of satellite communications \cite{Sanz2018,Fesquet2022}. Furthermore, QKD comes in two different flavors, DV- and CV-QKD. Due to the premature stage in which MW photon counters are, CV-QKD should be easier to implement with current MW technology \cite{Laudenbach}, although DV-QKD is expected to enable longer distance secure quantum key distribution.

In general, the motivations for CV-QKD are manifold. In traditional DV-QKD protocols such as BB84, one deals with single photons. Since there are no perfect single photons and single photon counters in the laboratories, and the security proof for BB84 relies on those, new protocols with relaxed security assumptions has been invented, such as measurement-device independent (MDI), device-independent (DI) and even more recently twin-field (TF) QKD. One may then argue whether it would make sense to invest on CV QKD, because we only rely on standard resources of the telecommunication industry: coherent states and (heterodyne) detection. In fact, there is much more interest from the telecom industry to develop coherent control than single photon states/detectors, as it would be useful also for classical telecommunications. 



Quantum key distribution with microwaves has not been realized yet. Seeing the performances obtained for time-bin generation \cite{Kurpiers2019} and photon counting (see Section \ref{subsec:detection}), it would be in principle possible to implement already the COW protocol with microwaves. As previously described, for microwave DV-QKD it is still early to use the polarization degree of freedom. Nevertheless, developments in this front would be extremely useful, as they would enable the use of more recent QKD protocols such as the three-state one-decoy BB84 \cite{Grunenfelder2018}. CV-QKD should be possible to realize today, as demonstrated by the recent publication \cite{Fesquet2022}. In this article, the authors show that using the protocol from Ref.~\cite{Cerf2001} based on Gaussian encoding of squeezed states, microwave CV-QKD can perform better under imperfect weather conditions.

\subsection{Scaling and integration}
\label{subsec:qLAN}

The realization of a scalable QLAN for quantum communication is an important milestone for the implementation of distributed quantum computing and for quantum internet applications \cite{Awschalom2021}. Such a network should provide an all-to-all connectivity between distant superconducting quantum processors and simultaneously maintain quantum coherence of interactions, including a possibility for entanglement distribution. There has been already significant efforts towards QLAN implementation in the optical domain at telecom frequencies over fiber networks and in the free-space environment. However, after years of research efforts, microwave-to-optics transduction experiments still have not reached sufficiently good efficiencies. The latter remain on the order of $10^{-5}$ in the single-photon regime \cite{Mirhosseini2020} after accounting for post-selection probabilities or added noise photons. Furthermore, the materials compatibility of such converters with that of high-performance superconducting qubits also remains to be an outstanding problem. In order to avoid these problems, we consider a direct realization of microwave quantum networks by connecting spatially separated dilution refrigerators via a cryogenic microwave link \cite{Deppe2020}. This approach has two important advantages. First, microwaves are the natural frequency scale of superconducting quantum circuits and, therefore, frequency conversion losses trivially vanish in this scenario. Second, since superconducting quantum processors require cooling to millikelvin temperatures, the cryogenic requirements are reduced to the development of a suitable millikelvin interface between dilution fridges containing distant superconducting circuits. We have implemented this goal within the EU quantum Flagship project QMiCS together with  Oxford Instruments Nanotechnology Ltd. (OINT). Our cryogenic link with a total length of 6.6\,m reaches temperatures below $\SI{35}{\milli \kelvin}$ and connects a home-built dry dilution refrigerator with a commercial Triton500 dilution refrigerator from OINT \cite{Batey2009}, both dilution refrigerators reaching below $\SI{20}{\milli \kelvin}$. During the design phase of the cryogenic link, several unique features were incorporated. One of these is a cold ‘junction box’ (a cold network node (CNN)) at the mid-point of the link. In addition, due to the large radiative heat load from such a system -- \textit{i.e.}, two dilution refrigerators connected by a long cryogenic link -- the cooling power at the higher temperature stages is important. This radiative heat load can be reduced by adding multi-layer insulation, but in order to add significant cooling power, an extra Pulse Tube Refrigerator (PTR) or 1 K pot can be added to the CNN. Other unique design features will be discussed in further publications \cite{Renger2022}. An inter-fridge quantum communication channel within the cryolink is realized with a superconducting Niobium-Titanium coaxial cable with characteristic losses around $0.001$\,dB/m. The CNN furthermore allows to connect additional link arms and, therefore, enables scaling to a two-dimensional quantum area network. All integral parts for such a system were supplied commercially by OINT and could be supplied for future projects. This ensures straightforward scalability to longer distances and allows near-future implementation of distributed superconducting quantum computing platforms operating in the microwave regime \cite{Krinner2019,Magnard2020b}.

\begin{figure*}[t!]
\includegraphics[width=.9\linewidth]{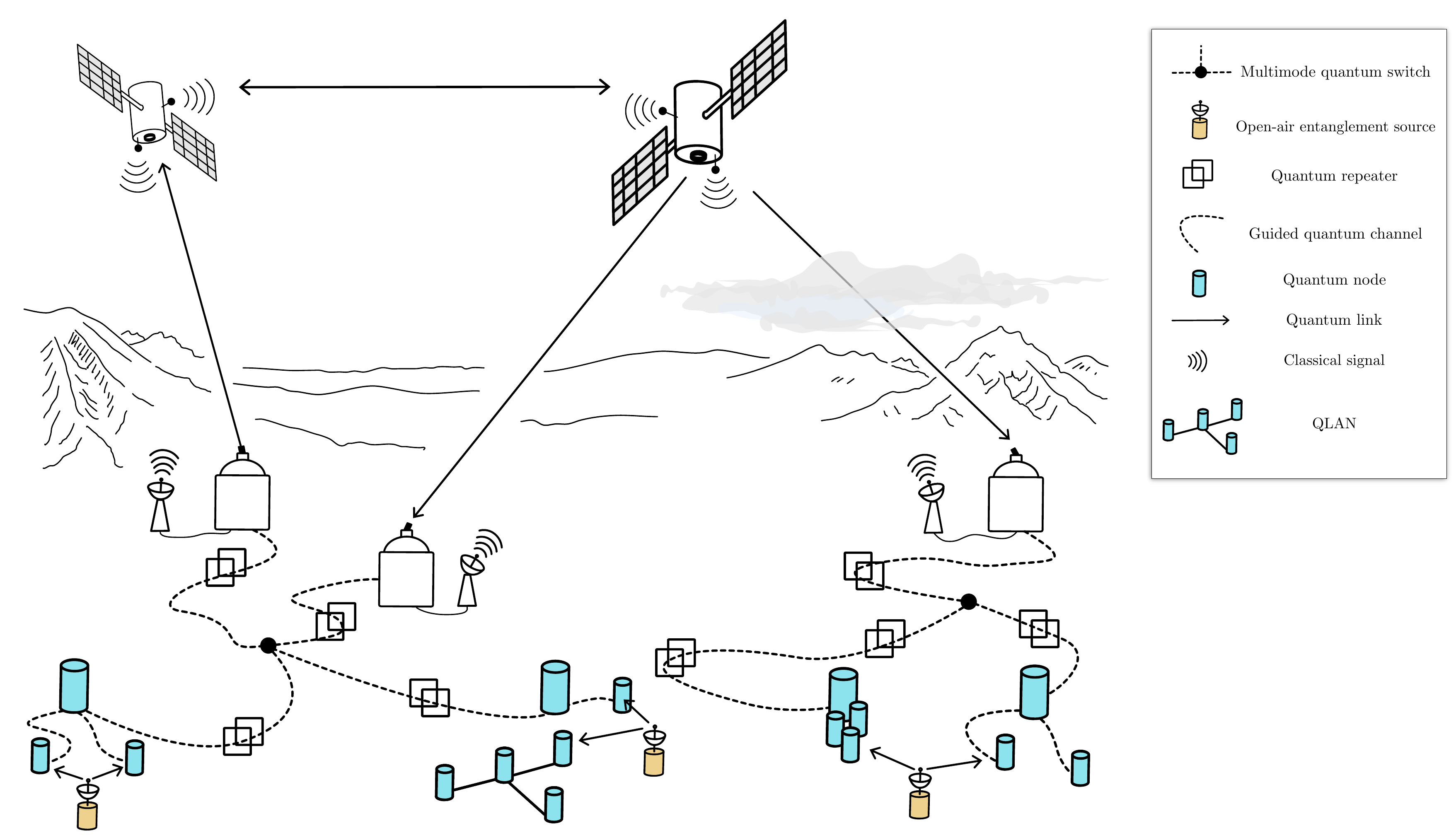}
\caption{Idealized view of a section of an interconnected, large-scale quantum communications network that combines optical and microwave links. Two low-orbit satellites capable of performing QKD  protocols are optically connected to distant Earth-based stations. Quantum repeaters are then used to distribute the quantum resources throughout an entire network, making use of trusted quantum switches to choose which node(s) receive the quantum information. Locally, open-air microwave entanglement sources are used to entangle remote but relatively close stations (up to \SI{200}{\meter}), making use of the benefits of microwaves in open-air transmissions with unfavorable weather conditions (see Refs.~\cite{GonzalezRaya2022, Fesquet2022}), and a simplified version of a QLAN, where cryolinks are used to connect different cryogenic stations to perform local quantum communication or sensing experiments.}
\label{fig:network}
\end{figure*}

A key, elusive element for the task of scaling-up a quantum communication network to cover large distances is the quantum repeater (QR). In Figure \ref{fig:network} we show an idealized quantum network that makes use of QRs for long-range quantum communications. Classically, a repeater, in its simplest form, is a device sitting between two nodes of a network that receives an incoming noisy signal, measures it, reproduces a noiseless copy of the signal, and sends it to the other node. Given the fact that quantum information can't be cloned, this notion of repeater has to be revised. In order to be compatible with all the communication advantages the quantum realm offers, the most notable is its intrinsic security, quantum tomography is not a solution. This is because, in short, a receive-measure-send scheme would destroy any possible initial entanglement between the two active nodes. To overcome this, three generations of QRs have been identified, depending on the techniques they use to improve the communication rate between nodes, in a sophistication-increasing order. The first generation uses heralded entanglement generation (\textit{e.g.}, via entanglement swapping), heralded entanglement purification/distillation, or a mix of both. Essentially, these repeaters rely on the ability to perform Bell measurements and two-way classical communication between the nodes, which already pose a challenge in microwaves, as argued in Section \ref{subsec:detection}. Additionally, Gaussian entanglement distillation requires many copies of the state, which may not be possible in some scenarios. The following proposals for QRs all apply to the CV case. The approach of noiseless linear amplification, weak measurements, and postselection of Ref.~\cite{diCandia2015} have been conceived specifically for microwaves. A QR based on quantum scissors and non-deterministic, non-Gaussian entanglement swapping was proposed in Ref.~\cite{Seshadreesan2020}. The entanglement distillation, and swapping techniques discussed in Ref.~\cite{GonzalezRaya2022} will also prove relevant for this task. Single-shot QRs will be required at some point, but this requirement most likely will imply the need for microwave quantum memories~\cite{Wu2010,Kubo2012,Saito2013,Yin2013,Wenner2013,Flurin2014,Bao2021,Palomaki2013,Palomaki2013a}, as explained in Ref.~\cite{dias2020}. Still, models for quantum communications without quantum memories have been proposed \cite{Munro2012}. The second and third generations of quantum repeaters use classical and quantum error correction, respectively, to solve additional sources of imperfection in the incoming signal, and are still beyond current reach, to the best of our knowledge. Further, in the specific case of propagating quantum microwaves, much work needs to be done, both theoretically and experimentally to produce operative quantum repeaters, even of first generation.


\section{Quantum Sensing}\label{QSensing}
Quantum sensing and metrology address a fundamental aspect of quantum mechanics
from a practical viewpoint: what is the maximum precision available at any given situation? Typically, quantum-enhanced sensing seeks an advantage provided by the resorting to quantum properties: entanglement, squeezing, or simply the use of single-photon states turn into \textit{resources} in this context. In this section we review some of the most important sensing protocols with potential applications in
quantum microwave technology.
\subsection{Quantum illumination}
Quantum illumination (QI) \cite{Lloyd1463, PhysRevLett.101.253601, guha2009,borre, shapiroStory, alsing2019} is a protocol that beats the classical signal-to-noise ratio (SNR) in the problem of detecting a low-reflectivity object embedded in a bright thermal bath by using signal-idler path entanglement as a resource. Importantly, QI preserves its quantum advantage even when the noisy, lossy channel is entanglement-breaking: The remaining quantum correlations (quantum discord) between the probe signal and the reference idler mode are enough to beat any classical scheme. Actually, the quantum discord directly quantifies the quantum advantage~\cite{Bradshaw2017}. 

In simple terms, the ideal case of Gaussian QI works as follows: $M \gg 1$ copies of a signal-idler symmetric two-mode squeezed vacuum state $\ket{\psi}_{SI}$ are prepared. The signal pulses, each containing $N_S$ photons on average, are sent to probe a given region at a given distance (both known) where there may be a low-reflectivity object, while the idlers are coherently kept in the lab. After the time associated with the assumed distance to the target, the receiver gets either just noise (the object is not there), or the reflected signals mixed with noise, in case the object is there. Assuming no bias in the presence/absence of the object, the problem reduces to a discrimination between two equally likely
states: one containing signals, one that doesn't. Theoretically, it is possible to obtain an enhancement of \SI{6}{\deci\bel} in the SNR, equivalent to an error probability four times smaller than any classical strategy. The full advantage of QI over the optimal classical approach using Gaussian states and detectors, namely the use of a coherent state as probe followed by any detection scheme using Gaussian operations and field quadratures measurements.
However, this \SI{6}{\deci\bel} enhancement requires a joint measurement between all signal and idler pairs, a very challenging task that essentially requires a quantum computer~\cite{zhuang2017,guha2009}. More modest approaches, which require only local operations and classical communication (LOCC), can obtain up to \SI{3}{\deci\bel} \cite{Sanz2017}, and could in principle be realised with efficient photon-counters.  

QI has been experimentally demonstrated in the optical regime \cite{genovese}. The main problem there is idler storage loss, which we address in this manuscript (Section \ref{subsec:qradar+imaging}). Experimental schemes for QI with microwaves have been proposed, but a fully-microwave implementation is  missing to the best of our knowledge: in Ref.~\cite{pirandolaMicrowave} the authors up-convert the signal to optical frequencies, with the corresponding conversion losses. In Ref.~\cite{shabir} a digital receiver is used, without any joint measurement and thus no quantum advantage. 

Still, recent developments in photon-counting techniques, such as qubit-based detectors proposed in Ref.~\cite{Dassonneville2020} or ultrasensitive bolometers developed in Ref.~\cite{Kokkoniemi2020}, could represent an important step towards the experimental realisation of QI in microwaves. Indeed, in Ref.~\cite{Sanz2017} it was shown that using a local approach based on linear operations and photon-counting the TMSV state obtains the \SI{3}{\deci\bel} gain mentioned above. A microwave-oriented study based on TMST states, and including decoherence in the idler mode due to imperfections in the delay line is needed in order to experimentally test these ideas. Another challenge for the realisation of microwave QI is the fact that the number of pulses $M = T W$, where $T$ is the duration of each signal-idler pulse and $W$ its phase-matching bandwidth, has to be very large ($M \sim 10^6$), because the theoretical advantage of QI is only guaranteed when the Chernoff bound is achieved, and this happens asymptotically with $M$. While this requirement is easily met by optical systems, it may be more challenging with microwaves.
However, the main caveat for a end-to-end microwave demonstration of QI remains: finding a way to perform a joint measurement on the signal and idler, which requires a measurement apparatus at low temperature. 

\subsection{Quantum radar and imaging}\label{subsec:qradar+imaging}
The \SIrange{1}{10}{\giga\hertz} transparency window of the atmosphere
has motivated for many years now the use of microwave frequencies for radar applications. The term `quantum radar' \cite{macconeRadar, lanzagorta} refers to a hypothetical device that would outperform any classical radar by resorting to quantum effects, typically entanglement. For some time, the QI protocol has been the main theoretical candidate for enabling such device, for two reasons: QI works best in the very low average signal photon number $N_S \ll 1$ and high noise regime $N_\text{th} \gg 1$. The first condition makes one think immediately of applying QI to a radar scheme, since radars typically want to detect without being detected; the second, points towards a  \textit{microwave} QI-based quantum radar because the atmosphere is naturally bright enough to meet $N_\text{th} \gg 1$ while being almost transparent: the expected number of photons at $T=300$ K and \SI{5}{\giga\hertz} is roughly 1250. So QI seems to be the perfect fit for enabling a quantum-enhanced radar, capable of obtaining an unprecedented detection precision, or even to `unveil' electromagnetically-cloaked objects, invisible to a classically-conceived apparatus \cite{Lasheras2017}. However, one must be careful before claiming that QI \textit{is} the way to go to construct a quantum radar solving the detection problem. First, a microwave QI-based quantum radar would need to solve the issue of idler storage loss:  just \SI{6}{\deci\bel} of loss in the idler delay line would automatically destroy the quantum advantage. The bound is twice as strict if the local approach providing \SI{3}{\deci\bel} of enhancement is used. Indeed, in Ref.~\cite{shabir} a bound of \SI{11.25}{\kilo\meter} on the maximum target distance was obtained, using an optical fiber delay line and up-conversion schemes. Second, as argued before, the pulse-time/bandwidth product requirement, $TW \gg 1$ appears very challenging with current microwave technology: as argued in Ref.~\cite{Sorelli2020}, the phase matching bandwidth for a microwave signal at \SI{10}{\giga \hertz} is $W \sim 100$ MHz, which gives a time per pulse of $T\sim10$ ms. Third, QI assumes that the location and velocity of the object is known. This translates in each pulse interrogating a single polarization-azimuth-elevation-range-Doppler region at a time. If the strategy is to be split into different bins, the quantum advantage decays logarithmically \cite{shapiroStory}. Still, $T\sim$ 10 ms is far too long to safely assume that the target has not moved. Fourth, even if all the above difficulties are circumvented,  microwave QI will still require cryogenics, which significantly increases the payload of the radar, something that may not be justified by the modest advantage of QI over classical target detection. A synthetic aperture quantum radar (SAqR) could find microwave imaging applications in high-orbit satellite-based stations, where the cooling power needed to achieve cryogenics is greatly reduced by the already cold temperature of space (2.7 K). We thus conclude that a commercial microwave quantum radar based solely on quantum illumination for target detection will not appear in the near future. Still, an inter-fridge experimental proposal for QI, where a TMST state is generated inside one of the cryostats, sending one mode through a cryolink equipped with controlled thermal noise, and interacting with a low-reflectivity target at the other cryostat, may allow to better test the limitations of QI in the microwave regime.
Other proposals for quantum radar, like the one in Ref.~\cite{macconeRadar} seem rather impractical in the microwave regime, due to the high sensitivity of the protocol to loss and noise. This is not the case of the Doppler quantum radar presented in Ref.~\cite{Reichert2022}. 
There, a signal beam is sent towards and reflected by a moving target, which shifts the signal frequency due to the Doppler effect. By measuring the frequency of the reflected signal, the radial velocity can be inferred. The protocol uses  signal-idler beams, with both frequency entanglement and squeezing as enhancing quantum resources. Contrary to QI, the protocol's quantum advantage grows with increasing signal power for most parameter regimes, that is, it beats the SQL and even reaches the Heisenberg limit for the majority of parameters. Another advantage over QI is that the idler beam does not need to be stored, which as we have argued can make QI problematic for real-life target detection in certain frequency intervals. The main results are for the lossless and noiseless case. However, a brief study of loss has shown that the protocol is fairly loss resilient: The lower bound for the quantum advantage in the presence of loss grows linearly with the degree of frequency entanglement, which is, in principle, unbounded. Thus, a quantum advantage could always be achieved for a high enough degree of frequency entanglement in a lossy quantum channel. The main challenge for a realisation of the protocol in the microwave regime is the implementation of the optimal measurement: a frequency-resolved photon counter for the signal and idler beams. Broad-band microwave photon-counters are thus required. The protocol can be adapted to range estimation, for which the optimal measurement is the arrival time of the individual signal and idler photons, which may be easier to implement for microwaves.
In this line, some recent work has focused on using QI for the ranging problem, with encouraging theoretical results of an improvement of $\mathcal{O}(10)$\,dB over the classical strategy \cite{Zhuang2021}. Despite all these important efforts, the gap between theory proposals and actual commercial devices is currently too big to expect operative microwave quantum radars in the near future.

`Quantum imaging' encompasses any technique benefiting from quantum effects -- such as entanglement, superposition or squeezing -- in order to enhance the contrast or resolution of an image. A notorious example is ghost imaging~\cite{simon2016}.
Biological tissue is mostly comprised of water, limiting microwave penetration depths to few cm. Shorter microwave wavelengths provide lower resolution widths, require smaller antennas, and face lower thermal background, at the cost of lower penetration depths.
To date, radar-like techniques have been successfully applied in diagnose imaging of breast cancer~\cite{fear2002, felicio2019}, benefiting from the low water content of breast fat and distinct dielectric properties between healthy and malignant tissue. 
Quantum advantages in the sensing of dielectric properties could enhance these protocols. A first proof of advantage in the computation of reflectivity gradients, using bi-frequency entangled probes, has been theoretically proposed~\cite{casariego2020}.
A QI-based medical imaging device avoids the first and third objections to the quantum radar in the previous paragraph, with idler storage times 5 orders of magnitude shorter (from km to cm), and a static target. Nevertheless, the high in-tissue attenuation and corresponding low penetration depth is a new obstacle to overcome. 
A transmission image of an object with transparent-opaque contrasts has been obtained with QI, achieving an advantage in the rejection of background noise, in optical frequencies~\cite{gregory}. In principle, this protocol could be adapted to microwave frequencies.

In general, we expect that taking advantage of the quantumness of microwaves in order to perform a quantum imaging protocol will only be justified in systems that can sustain cryogenic temperatures. As an example, recent developments in microwave photon-counting techniques could have an impact in magnetic resonance experimental sensitivities, like the effort of resorting to squeezed states in order to perform electron spin resonance spectroscopy~\cite{Bienfait2017}.

\subsection{Inference of quantum system-environment interactions}


Advances in propagating quantum microwaves could be further boosted by the sensing of the way the external environment dynamically affects nominal working conditions. In other words, one would ask for sensing protocols able to infer quantum system-environment interactions. In this context, one of the most promising applications is quantum thermometry, that may play a crucial role to improve accuracy in carrying out communication and computing in the quantum regime~\cite{KurpiersPRApp2019,SultanovAPL2021,DanilinArXiv2021}. It is generally known, indeed, that the unavoidable influence of the environment can lead both to a temperature increase and energy fluctuations in terms of heat losses~\cite{ManzanoPRX2018,BatalhaoChapter}, which is expected to prevent the correct functionality of the physical mechanism one is investigating. In this regard, it is worth mentioning recent studies about irreversible losses in quantum logic gates~\cite{CimininpjQI2020} and quantum annealers~\cite{GardasSciRep2018,BuffoniQST2020}. Of course, a similar evidence also holds if one deals with quantum microwaves. In fact, one may consider an optical quantum memory~\cite{Morton2008,Simon2010} in solid state that is a perfect example of a quantum system, based on microwave transitions, subject to environmental degrees of freedom. A quantum memory usually consists of a $\Lambda$-system \footnote{A generalized  $n$-level $\Lambda$-system is  described in the rotating frame by a Hamiltonian of the form $H=\sum_{k=1}^n \Omega_k \ket{e}\bra{k} + \Omega_k^* \ket{k}\bra{e}$, where $\lbrace \Omega_k \rbrace_k$ are the Rabi frequencies associated with the transition from the $k$-th level $\ket{k}$ to the excited state $\ket{e}$. The name $\Lambda$-system comes from the way the levels are arranged in the simplest case of $n=2$.} with a microwave transition that is used for quantum storage. For its functionality, one needs to reduce the effect of decoherence on the spin systems that compose the memory. Such decoherence mechanisms can have different origins; among the most relevant we recall the interaction between magnetic dipoles, and the coupling of the spins to phonons. In general, for solid state media, dynamical decoupling enhanced magnetometry~\cite{Pham2012,Baumgart2016} already provides information about locally varying magnetic fields causing decoherence. In such a framework, we are confident that decoherence and relaxation processes can be generally inferred by means of quantum sensing protocols. From this point of view, quantum thermometry itself can be seen effectively as a branch of quantum sensing in the sense that one aims at deciding whether the quantum system of interest is in contact, or not, with the external environment that leads to thermal fluctuations.

With quantum propagating microwaves in mind, we propose a quantum estimation strategy with a direct application to thermometry, by taking inspiration from recent results in Refs.~\cite{RossiPRL2020,MontenegroPRR2020,SongPRA2021}. Our proposal is based on using a cavity-system that is resonantly driven by a coherent laser pulse. The quantum system that one aims to investigate is inserted in the cavity. The system is assumed to be in contact with an external environment exhibiting macroscopic features, as for example a thermal bath. The interaction between the quantum system and the environment leads to fluctuations of physical quantities, such as energy and/or temperature. Here, it is worth observing that the assumption of considering the presence of the environment (within the open quantum system framework) stems from our impossibility to isolate the dynamics of the quantum process of interest. One of the main reasons for that is the inaccessibility of the latter, which may be physically embedded within a body with larger dimension, \textit{e.g.}, a mesoscopic system. Concrete examples are actually provided by quantum systems with microwave transitions, as rare-earth ions in solids~\cite{Tittel2010,ZambriniCruzeiro2017}, nitrogen vacancy (NV) centers~\cite{Shim2013,Heshami2014}, silicon vacancy centers~\cite{Sukachev2017}, or even molecules~\cite{Rabl2006}. 

For all these quantum systems, the scope of the cavity-based setup is to scan -- and possibly reconstruct -- the way the system is in contact with the environment, and how it leads the system to relax towards a steady-state. The thermalization towards a steady-state with a well-defined temperature is a special case of the latter. We recall that in the case of systems with microwave transitions in solid media, this kind of sensing strategy is expected to be useful in studying the spin lattice relaxation process~\cite{Orbach1961}, which is important to improve quantum information processing devices like quantum memories~\cite{ZambriniCruzeiro2017}. Specifically, the scanning using a cavity-system is enabled by performing a continuous sequence of weak measurements, without completely destroying coherence in the measurement basis~\cite{WallraffNature2004,MurchNature2013,TanPRL2015}. The output field from the cavity is then physically monitored by means of a detector that depends on the specific quantum measurement observable. For example, in~\cite{RossiPRL2020} a homodyne receiver is employed, while in~\cite{SongPRA2021} energy projective measurements are performed. In the context of propagating quantum microwaves, it is already known that homodyne detection can be employed. However, as pointed out in Ref.~\cite{TycJPAMG2004}, also a projective measurement of energy might be implemented, thanks to the evidence that a balanced homodyne detection, under specific conditions, can reproduce the effect of performing projective measurements of the quadrature phase of the output signal field (\textit{i.e.}, CVs Bell measurements). Furthermore, one could be also interested in addressing quantum sensing in non-Gaussian regimes, for which non-Gaussian measurements are required. For such a task, homodyne detection has to be replaced or supplemented by photon-counting~\cite{QiOE2020}; in this context, the photon-counting device in~\cite{Wang2019}, for the sensing of microwave radiation at the sub-unit-photon level, could be a strong candidate.

Overall, the proposed method based on continuous quantum measurements is expected to efficiently acquire information about the environment to which the analyzed system is in contact (or in which it is embedded) by making use of a sequence of outcomes that are continuously recorded by the monitoring process. Such an information is granted, at the price of taking into account both the quantum measurement back-action (below denoted as ``mba'') and the stochastic contribution (``stoc'') to the dynamics induced from conditioning upon the measurement records~\cite{TanPRL2015,RossiPRL2020,SongPRA2021}, described respectively by the super-operators $\mathcal{L}_{\rm mba}$ and $\mathcal{L}_{\rm stoc}$. Refer to Refs.~\cite{TanPRL2015,RossiPRL2020} for a detailed description of the latter that holds independently on the system one monitors (in our case, a mechanism related to propagating quantum microwaves as for example the one detailed below when discussing the system operator $\hat{A}$). Hence, formally one has to consider the conditional dynamics of the quantum state $\rho$, governed by a \emph{stochastic master equation} of the form 
\begin{equation}\label{eq:stoc_master_eq}
    \diff \rho = \left(\mathcal{L}_{\rm bath} + \mathcal{L}_{\rm mba} + \mathcal{L}_{\rm stoc}\right)\rho \, \diff t \,. 
\end{equation}
In Eq.~(\ref{eq:stoc_master_eq}), $\mathcal{L}_{\rm bath}$ denotes the super-operator modelling the interaction between the quantum system and the bath (with macroscopic features) in its surrounding, since for the sake of simplicity we are here considering the particular case that the environment is a thermal bath. A common expression for $\mathcal{L}_{\rm bath}\rho$ is the following:
\begin{equation}\label{eq:L_bath}
\mathcal{L}_{\rm bath}\rho = \Gamma \left(n_{\rm bath} + 1\right)\mathcal{D}[\hat{A}]\rho + \Gamma\, n_{\rm bath} \mathcal{D}[\hat{A}^{\dag}]\rho 
\end{equation}
where $\Gamma$ denotes the energy damping rate and $n_{\rm bath}$ is the bath occupancy. Moreover, in Eq.~(\ref{eq:L_bath}), $\mathcal{D}[\hat{A}]\rho \equiv \hat{A}\rho\hat{A}^{\dag} - \{\hat{A}^{\dag}\hat{A},\rho\}/2$ (usual super-operator in Lindblad form with $\{\cdot,\cdot\}$ denoting the anti-commutator), and $\hat{A}$ is the quantum system operator on which the bath acts. A concrete example for the operator $\hat{A}$, involving quantum microwaves, can be found in Ref.~\cite{NorambuenaPRB2018} where the spin-lattice relaxation of individual solid-state spins in diamond NV centers is studied. In fact, it is there shown that dissipative spin-lattice dynamics -- induced both by phonon interactions and noise due to magnetic impurities -- act isotropically on the angular momentum operators that rule the spin transitions along each space coordinates. Thus, once known the system-bath interactions (even on average), the goal of the sensing strategy we are here proposing is to extract information about $\Gamma$ and $n_{\rm bath}$ from the weak measurement record $r(t)$ (time-varying signal) that is measured as output field of the cavity-system. A possible expression for $r(t)$, valid also if we deal with quantum microwaves, is $r(t)={\rm Tr}\left[\rho(t)\sigma_z\right] + \diff\mathcal{W}$ with $\diff\mathcal{W}$ denoting the zero-mean Gaussian distributed Wiener increment and $\sigma_z$ the Pauli matrix $Z$. This procedure, which well fulfills with the present roadmap, will be the subject of a forthcoming research activity. 

To conclude, we stress that this formalism can take into account also fluctuation terms, since it is able to directly scan an open quantum dynamics at the single-trajectory level. Hence, by means of such a high degree of resolution, it is expected to provide the sufficient amount of information to: \textit{(i)} infer how a quantum system physically interacts with the external environment, and \textit{(ii)} reconstruct some key bath parameters, as for example $\Gamma$ and $n_{\rm th}$ in the model of Eq.\,(\ref{eq:stoc_master_eq}). Finally, notice also that the proposed method works for both discrete and continuous variable quantum systems. 

\subsection{Direct Dark Matter detection}\label{subsec:DarkMatter}
To illustrate the reach of microwave quantum sensing, we close with a discussion of novel ways to look for 
dark matter candidates.
Dark matter is predicted to make up to 85\% of the matter in the Universe, based on gravitational observations compatible with General Relativity. If interpreted within the context of particle physics theories, these imply a wide range of possible masses for dark matter particles. For example, currently many new technologies are pursued in order to search for Weakly Interacting Massive Particles (WIMPs) with masses in the meV-GeV range; for a recent review, see Ref.~\cite{Essig2022}.

Within the context of this paper, two especially interesting dark matter candidates are axion like particles and dark photons. The former is originally motivated as a solution to the strong CP problem in the Standard Model of elementary particle interactions, while the latter generally arises when extending the Standard Model with a new U(1) gauge symmetry. 

The axion is predicted to have a very weak coupling with electromagnetism, and this allows to search for a weak narrow band signal at a frequency corresponding to the unknown mass of the axion in cavity based detectors known as haloscopes. The dark photon, on the other hand, is constrained by observation to have only extremely weak coupling with electromagnetism and axion haloscopes can therefore be used to search for dark photons, as well. Haloscopes have been used to search for axionic dark matter in the \SIrange{1.8}{24}{\micro e\volt} mass range~\cite{Brubaker2017,Semertzidis2022,Chadha-Day2022}, and the results have also been applied to constrain dark photons in this mass range~\cite{Ghosh:2021ard}.

This method, however, is mostly hindered by the fact that these searches can be slow and that it is difficult to overcome the standard quantum limit, where cavity measurements introduce noise due to quantum uncertainty.

Several improvements for this technique are now being considered. These may bring relevant results not only for axion searches,
but also for general metrology beyond the standard quantum limit.
Backes et al.~\cite{Backes2021} introduce quantum enhancements to the HAYSTAC, which allows for a small speed improvement in the search
and exploration beyond the quantum limit with the microwave-frequency field prepared in a squeezed state. Wurtz et al.~\cite{Wurtz2021} suggest and implement a way of amplifying the axion signal through mode squeezing and state swapping interactions before any noise from the quantum limit actually contaminates the signal. 
Dixit et al.~\cite{Dixit2021} introduce a new method for cavity searches which circumvents the quantum uncertainty generated by performing measurements in
the cavity by instead performing a photon counting technique which does not destroy the photon. This is possible with the use of a superconducting qubit, as discussed above, and they demonstrate the technique for a dark photon search.
Recent developments in ultrasensitive bolometers \cite{lowNEP,Kokkoniemi2020, Lee2020} and calorimeters \cite{Karimi2020} show promise in overcoming the quantum noise in the cavity measurements especially in the higher mass end \cite{Lamoreaux2012}, as they have demonstrated sensitivities with potential for single photon detection down to tens of GHz frequencies. These detectors are also particularly appealing candidates for experiments that rely on photon-counting due to their ability to absorb radiation from a wide band. A plan for a new generation of haloscope experiments employing thermal detectors for axion search in the mass range of $10^{-3}$--1~\SI{}{e\volt} has been proposed~\cite{LiuAxion2021}. 

The profound observable for precision cosmology is the cosmic microwave background (CMB). The main focus of current observational efforts is in the so called polarization B-modes, as their observation would provide evidence for the exponential expansion (inflation) of the very early Universe. Bolometric interferometry to measure these modes has been proposed in several projects currently combined in the QUBIC experiment~\cite{QUBIC:2020kvy}. Precision measurements of the CMB may also be relevant for axion searches: oscillations of the axion field have recently been predicted to lead to two different effects in the polarization of the CMB \cite{Fedderke2019}. A uniform reduction of the polarization is predicted to result from multiple oscillations of the axion field occuring during the CMB decoupling epoch in the early universe. On the other hand, present day oscillations of the axion field are predicted to lead to a real-time AC oscillation of the polarization of the CMB. The authors of \cite{Fedderke2019} state that observing the latter effect would be a particularly convincing evidence for the existence of axions in the lowest mass range, but these experiments require dedicated time-series analysis of the CMB signal. Compared with the detectors used in the previous CMB experiments \cite{Pirro2017}, bolometers with improved sensitivities may help in faster mapping of the CMB background across the sky.

\section{Conclusions}
 \begin{figure*}[t!]
\includegraphics[width=\linewidth]{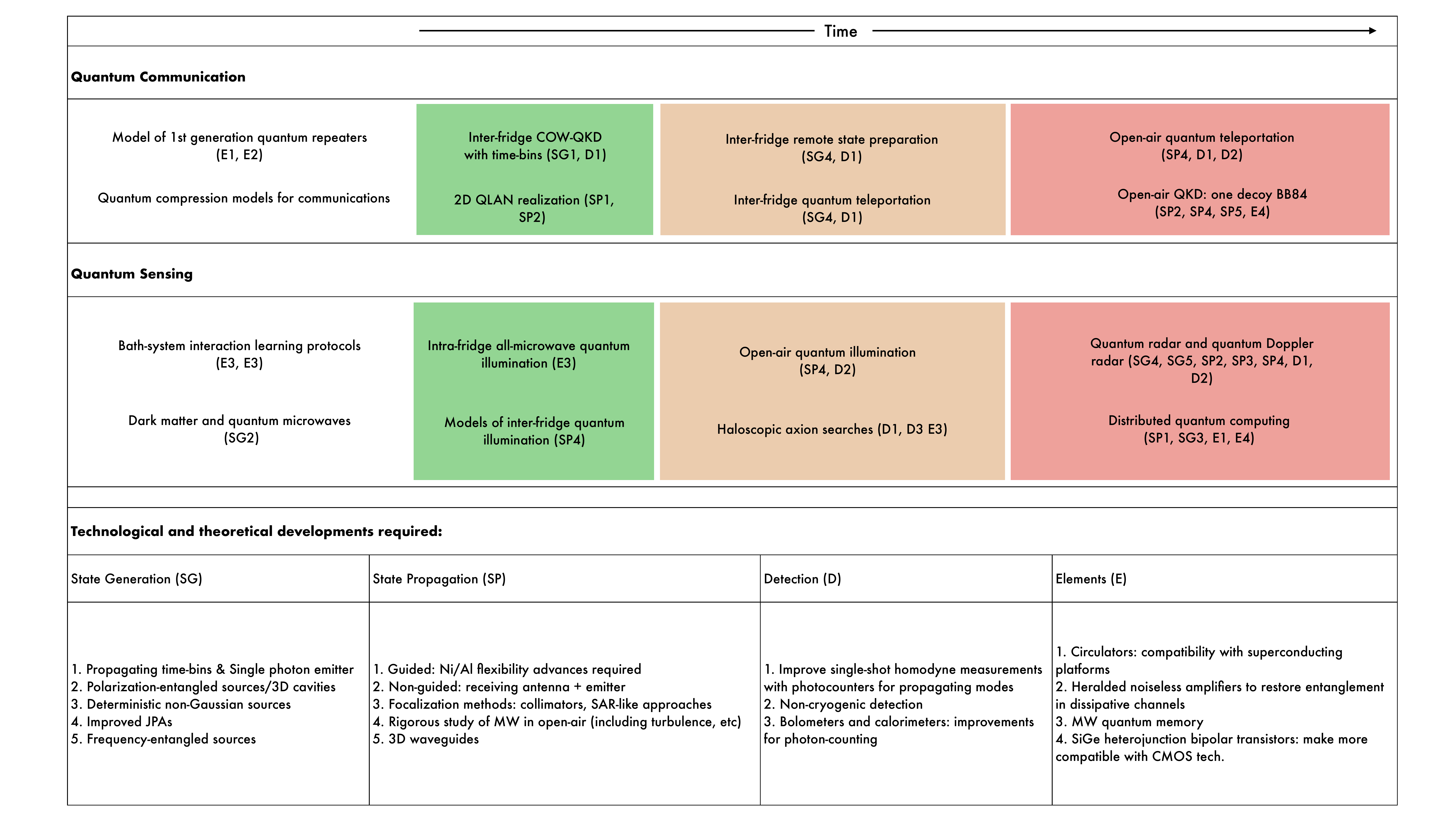}
\caption{Quantum communication and sensing roadmap for the upcoming years. The first part of the table consists in the quantum communication and sensing protocols and results we think should be addressed in the mid-term future. The first box on the left contains proposals for experiments that are typically on an early stage, for example the quantum microwaves-dark matter detection interplay we discussed in Section \ref{subsec:DarkMatter}. Then, three coloured boxes contain experimental proposals with an increasing difficulty factor to the right. In green, we have experiments that can already be performed, with very minor adjustments to current technology; in orange, we include experiments that should become proof-of-principle results for future real-life technologies and applications of quantum microwaves. In red, finally, we have included some of the necessary results discussed along this manuscript, but that we think will require years. Importantly, the second part of the table consists on technical results, roughly divided by quantum-mechanical categories, that will enable the technologies described before. To link the two tables, we have included which are the requirements of each of the quantum communication and sensing experiments, hoping to compress the ideas and to trigger motivation for doing more research in the field.}
\label{fig:roadmap}
\end{figure*}

In this manuscript we have addressed the challenges that the field of propagating quantum microwaves faces in order to find real-life applications in quantum communications and sensing. To this end, we started with a state-of-the-art description of the essential stages that are a common denominator for both communications and sensing protocols, such as state generation (including entanglement rates), guided and non-guided propagation, amplification, beam splitting, detection schemes and photon-counting devices, among others. Along the way, we pointed the aspects that require further research, both from a theoretical and the experimental point of view. Then, we discussed quantum communications in length, with a focus on the need for both continuous variables and discrete variables implementations of microwave quantum key distribution, for which the existence of an operative quantum local area network (QLAN) making use of cryolinks to connect remote cryogenic refrigerators will represent a major step towards proof-of-principle experiments. A discussion on the scalability of such QLANs and their integration within the quantum internet followed, concluding that both an efficient microwave-to-optical, and a fully-microwave platforms are needed and expected to coexist. Then we moved on to discuss quantum sensing, where quantum illumination and its relation to quantum radar were treated. Here, the conclusions were, first, that a fully-microwave demonstration of quantum illumination is still missing but the recent advances in photon-counters for propagating microwaves will surely solve this, and second, that a quantum radar based solely on this protocol is still far of reach. A brief discussion of some areas where quantum imaging protocols may benefit from the new results in photon-counting followed, though only in scenarios where the system can sustain a cryogenic temperature, ruling out medical imaging. Finally, we proposed two novel ideas in two fields that are wildly apart: thermometry or, more generally, the inference and characterization of system-environment interactions, and axionic dark matter search. In the first, we argued that a continuous sequence of weak measurements making use of the recent advances in both single-shot homodyne detection and photon-counting could lead to an efficient method for the quantum estimation of parameters such as temperature, or the expected number of photons present in a thermal bath. As for axionic dark matter, we investigated two novel ways to infer the existence of the elusive particle: one where the axion may generate  detectable microwave photons, and other where its interaction with the cosmic microwave background leads to two measurable effects in the polarization. 

As a closing remark, we stress what are in our view the most urgent needs for the future of applicable quantum microwave technologies: to solve the impedance-matching problem that the open-air transmission poses in order to design quantum-capable emission and reception antennae; to rigorously investigate the quantum-channel capacity of the atmosphere from a quantum-theoretic perspective, in order to identify the scenarios where quantum microwaves are expected to beat the telecom frequencies (visible and near IR), and taking into account the high entanglement production rates that come from the strong interactions of microwaves with non-linear devices; to make a proof-of-principle demonstration of a QLAN with operative superconducting chips for distributed quantum computing; and to realistically assess the validity and usefulness of microwave quantum illumination for quantum radar by making an open-air experimental proposal. To summarize, in Figure~\ref{fig:roadmap} we have condensed the results of the manuscript in a roadmap fashion, hoping to give a bigger picture and to trigger new ideas and research in this fascinating field of propagating quantum microwaves.

\acknowledgements
The authors thank the support from project QMiCS (820505) of the EU Flagship on Quantum Technologies. MC, EZC, RA, GF, SG, and YO thank the support from Funda\c{c}\~{a}o para a Ci\^{e}ncia e a Tecnologia (Portugal), namely through project UIDB/50008/2020, as well as from project TheBlinQC supported by the EU H2020 QuantERA ERA-NET Cofund in Quantum Technologies and by FCT (QuantERA/0001/2017). MC thanks C. Stolhe for useful discussions, and acknowledges support from the DP-PMI and FCT through scholarship PD/BD/135186/2017. GF acknowledges support from FCT through scholarship SFRH/BD/145572/2019. TG-R, MR and MS acknowledge financial support from Basque Government QUANTEK project from ELKARTEK program (KK-2021/00070), Spanish Ram\'{o}n y Cajal Grant RYC-2020-030503-I, as well as from OpenSuperQ (820363) of the EU Flagship on Quantum Technologies, and the EU FET-Open projects Quromorphic (828826) and EPIQUS (899368), and IQM Quantum Computers under the project “Generating quantum algorithms and quantum processor optimization”. MR acknowledges support from UPV/EHU PhD Grant PIF21/289. MM and GC acknowledge funding from the European Research Council under Consolidator Grant No. 681311 (QUESS), and from the Academy of Finland through its Centers of Excellence Program (project Nos. 312300 and 336810) and QEMP project (319579). 
MR, KGF, and FD acknowledge support by the German Research Foundation via Germany’s Excellence Strategy (EXC-2111-390814868), the Elite Network of Bavaria through the program ExQM, the EU Flagship project QMiCS (Grant No. 820505), the German Federal Ministry of Education and Research via the projects QUARATE (Grant No. 13N15380) and QuaMToMe (Grant No. 16KISQ036), and the State of Bavaria via the Munich Quantum Valley and the Hightech Agenda Bayern Plus.

\appendix


\end{document}